\documentclass[letter,fleqn,usenatbib]{mnras}

\usepackage{newtxtext,newtxmath}

\usepackage[T1]{fontenc}
\usepackage{ae,aecompl}

\usepackage{euscript}

\usepackage{graphicx}	
\usepackage{amsmath}	

\title[Proper motions of isolated Local Group galaxies]{Solo dwarfs III: Exploring the orbital
  origins of isolated Local Group galaxies with Gaia Data Release 2}

\author[Alan W. McConnachie et al.]{Alan W. McConnachie$^{1}$\thanks{E-mail: alan.mcconnachie@nrc-cnrc.gc.ca},
Clare R. Higgs$^{2}$, Guillaume F. Thomas$^{3}$, Kim A. Venn$^{2}$, Patrick
C{\^o}t{\'e}$^{1}$, \newauthor Giuseppina Battaglia$^{3, 4}$, Geraint
F. Lewis$^{5}$.
\\
$^{1}$NRC Herzberg Astronomy \& Astrophysics, 5071 West Saanich Road,
Victoria, British Columbia, Canada V9E 2E7\\
$^{2}$Physics \& Astronomy Department, University of Victoria, 3800 Finnerty Rd, Victoria, B.C., Canada, V8P 5C2\\
$^{3}$Instituto de Astrofísica de Canarias, Calle Via Láctea s/n, E-38206 La Laguna, Tenerife, Spain\\
$^{4}$Universidad de La Laguna, Avda. Astrofísico Fco. Sánchez, La Laguna, 38200 Tenerife, Spain\\
$^{5}$Sydney Institute for Astronomy, School of Physics, A28, The University of Sydney, NSW 2006, Australia\\
}

\date{Accepted XXX. Received YYY; in original form ZZZ}

\pubyear{2020}

\begin{document}
\label{firstpage}
\pagerange{\pageref{firstpage}--\pageref{lastpage}}
\maketitle

\begin{abstract}
  We measure systemic proper motions for distant dwarf galaxies in the
  Local Group and investigate if these isolated galaxies have ever had
  an interaction with the Milky Way or M31. We cross-match photometry
  of isolated, star forming, dwarf galaxies in the Local Group, taken
  as part of the {\it Solo} survey, with astrometric measurements from
  Gaia Data Release 2. We find that NGC 6822, Leo A, IC 1613 and WLM
  have sufficient supergiants with reliable astrometry to derive
  proper motions. An additional three galaxies (Leo T, Eridanus 2 and
  Phoenix) are close enough that their proper motions have already
  been derived using red giant branch stars. Systematic errors in Gaia
  DR2 are significant for NGC 6822, IC 1613 and WLM. We explore the
  orbits for these galaxies, and conclude that Phoenix, Leo A and WLM
  are unlikely to have interacted with the Milky Way or M31, unless
  these large galaxies are very massive ($\gtrsim 1.6 \times 10^{12}\,M_\odot$). We rule out a past
  interaction of NGC 6822 with M31 at $\sim 99.99\%$ confidence, and
  find there is a $<10$\% chance that NGC 6822 has had an
  interaction with the Milky Way. We examine the likely origins of NGC
  6822 in the periphery of the young Local Group, and note that a
  future interaction of NGC 6822 with the Milky Way or M31 in the next
  4\,Gyrs is essentially ruled out. Our measurements indicate that
  future Gaia data releases will provide good constraints on the
  interaction history for the majority of these galaxies.
\end{abstract}




\section{Introduction}

The isolated dwarf galaxies of the Local Group are the ideal systems
for understanding the galactic-scale consequences of physical
distancing. There seems little doubt that these dwarfs are not
currently exposed to the level of interactions that are relatively
common-place when orbiting in the more densely populated satellite
systems of the Milky Way and M31. What is unclear, however, is if all
these galaxies have forever been as isolated as they are now, or if an
interaction early in the life of the dwarf caused it to travel
to the outskirts of the Local Group in which it is now found.

While there are approximately 100 dwarf galaxies known within the Local
Group, most are satellites of the Milky Way or M31. Only about a dozen
dwarfs are not clearly associated with one of these two systems,
insofar as they are found more than $\sim 300$\,kpc (the approximate
virial radii of the halos of these two galaxies, e.g.,
\citealt{klypin2002, posti2019}) from both these
large galaxies. It seems unlikely that these dozen galaxies are the
totality of all isolated galaxies in the Local Group, and selection
effects are likely important. While there has been a concerted efforts
to find new satellites of the Milky Way and M31 in recent
years (for a list of Milky Way discoveries, see
\citealt{drlicawagner2020} and references therein; for a list of M31
discoveries, see \citealt{mcconnachie2018c} and references therein), the same is not true for more isolated Local
Group members. Such searches are difficult because of
their large distances (rendering member stars relatively faint in comparison to
Milky Way satellites) and the fact that they could be found anywhere
in the sky (in comparison to the
relatively small area of
sky around M31 in which its satellites are found). To the best of our
knowledge, the 
search for new Local Group dwarf galaxies using the Palomar Observatory
Sky Survey and ESO/Science Research Council survey plates by
A. Whiting and colleagues remains the
most comprehensive search to-date for which completeness estimates are
available. This search led to two new galaxies (Antlia and Cetus) out
of 206 candidates (\citealt{whiting1997, whiting1999}). \cite{whiting2007} estimate that their by-eye
searches for Local Group dwarfs are essentially complete down to a surface brightness of
$26 - 27$\,mags\,arcsec$^{-2}$ away from the plane of the Milky Way. This surface
brightness is comparable to that of Sextans, the lowest surface
brightness ``classical'' satellite of the Milky Way. In this
case, Leo T (\citealt{irwin2007}) - at a surface brightness of around
25 \,mags\,arcsec$^{-2}$ - would be one of only a few galaxies at these
surface brightnesses that was not already identified. Clearly, a new
search for distant Local Group galaxies (not satellites of M31 and the
Milky Way), using the extensive digital sky surveys
accumulated since the days of the Palomar Sky Survey, is overdue.

Isolated dwarf galaxies in the Local Group have long been noted to
exhibit some general differences to the satellite populations
(\citealt{einasto1974}). The former generally host younger, bluer,
populations, and have significant gas fractions. The latter generally
lack significant young populations and have low or negligible gas
fractions. Many exceptions to this trend exist (e.g, the LMC, SMC,
M33, Cetus, Tucana, Andromeda XVIII). Detailed star formation
histories of galaxies throughout the Local Group (e.g.,
\citealt{weisz2014} and references therein) suggest that in large part
the star formation histories of both satellites and isolated systems
are generally similar over most of cosmic time, and that their star
formation histories differ most significantly only in the last
gigayear or so. Thus, much of the difference in these populations
connects back to the fact that the isolated dwarfs have generally been
able to retain their gas to the present day, whereas many satellites
have not.

The position–morphology relation that has been identified in the Local
Group is seen elsewhere in the Universe (e.g.,
\citealt{bouchard2009, geha2012}), and prompts consideration of the influence of
nearby large galaxies on the evolutionary paths of dwarfs. Mechanisms
at play may include ram pressure stripping (e.g.,
\citealt{mori2000, vollmer2000, boselli2014}), tidal stripping (e.g.,
\citealt{penarrubia2008b}), tidal stirring (e.g., \citealt{mayer2006, kazantzidis2011}),
  induced star formation, and/or strangulation (\citealt{kawata2008}
  and references therein). Comparison of the properties of isolated
  dwarfs to satellites could therefore shed light on the importance
  and interplay of  several complex physical
  processes.

 An essential element of these considerations is the acquisition of
  comparable datasets for isolated dwarfs as exist for satellite dwarfs in
  the Local Group. It is with this in mind that we constructed the
  {\it Solitary Local (Solo) Dwarfs} survey, where our intent is to
  provide homogeneous, high-quality, wide-field optical
  characterization of the closest, isolated dwarf galaxies (see \citealt{higgs2016},
  hereafter Paper I). These dwarfs have been identified based on their
  current locations, namely that they are more than 300\,kpc from either
  the Milky Way or M31, and are within 3\,Mpc. The full sample
  includes, but is not limited to, the Local Group. For that subset
  that are within the Local Group, an important consideration is whether or not the present
  positions of these dwarfs are enough to indicate that they have
  always been isolated. 

  There are some dwarf galaxies in the Local Group, such as DDO210 or
  the Sagittarius dwarf irregular galaxy (Sag DIG), that are more than
  a megaparsec from either of the dominant two-some. In these extreme
  cases, then barring pathological velocities (which are not implied
  by their heliocentric radial velocities), there is simply not enough
  time in the Universe for these galaxies to have previously
  interacted with M31 or the Milky Way, and then to have reached their
  current positions (e.g., see Figure~8 of \citealt{mcconnachie2012}). But for systems that are of order 300\,kpc -- 1\,Mpc
  distant from the large galaxies, it is quite conceivable that they
  have had a previous interaction (and so potentially been subject to
  a whole gamut of complex environmental processes), and
  have since travelled to apocenter far from their
  host.

  Identification of these so-called backsplash systems is
  important since, unlike truly isolated systems, their properties
  cannot necessarily be attributed only to secular processes. Early numerical work by \cite{gill2005} on backsplash galaxies
  concluded that half of the dwarfs at $1 - 2 R_{vir}$ from an
  $L_\star$ galaxy could actually be backsplash systems. More
  recent work by \cite{teyssier2012} and \cite{buck2019} reinforce this finding, and attempt
  to identify which Local Group dwarfs are most likely backsplash
  systems based on their positions and/or radial velocities and/or velocity
  dispersions. Both studies conclude that there is more than a 50\% of
  NGC6822 and Leo T having passed close to the Milky Way. Of the remaining isolated Local
  Group galaxies, Andromeda XXVIII, Cetus, Eridanus 2, IC 1613,
  Pegasus DIG, Phoenix and Tucana are highlighted as possible backsplash
  systems (more than a 50\% chance) by one of the studies.

  Determination of a galaxy's orbit is the most robust method to
  identify if it is a backsplash system. Observationally, this
  requires its position on the sky, its distance, its radial velocity
  and its proper motion. The first three are all relatively simple
  measurements. The fourth is not. The first proper motion of a Local Group galaxy that
  was not in the Milky Way subgroup was made 15 years
  ago. \cite{brunthaler2005} identified a water maser in M33, and were
  able to use the Very Large Baseline Interferometer (VLBI) to obtain
  a systemic proper motion after correction for M33's rotation. Water
  masers are only associated with intense star formation, and the most
  intensely star forming galaxy in the Local Group is the dwarf
  starburst galaxy IC10. \cite{brunthaler2007} were able to measure its proper
  motion using an identical methodology as for M33. M31 itself is a relatively quiescent
  galaxy in comparison to these two satellites (e.g.,
  \citealt{davidge2012a} and references therein), and its proper
  motion has since been derived from
  observations with the Hubble Space Telescope (HST;
  \citealt{sohn2012}). Proper motions have also just recently been
  measured by HST for NGC147
  and 185 (\citealt{sohn2020}).

The advent of Gaia Data Release 2 (DR2; \citealt{gaia2018b}) has been a watershed
  moment for dynamical studies of the Milky Way satellite system, with
  a large number of papers dedicated to the measurement of almost all the dwarf galaxy
  satellites of the Milky Way (\citealt{helmi2018b, massari2018,simon2018, simon2020, fritz2018,
  fritz2019,kallivayalil2018, pace2019, longeard2018, longeard2020a, longeard2020b,
  mau2020, mcconnachie2020}). This sample includes three galaxies - Phoenix,
  Eridanus II and Leo T - that are in the {\it Solo} sample (all located at
  $\lesssim 450$\,kpc).  In addition, Gaia DR2 has been used to
  independently confirm the previous measurements of the proper
  motions of M31 and M33 (\citealt{vandermarel2019}), albeit with less
  accuracy than HST or VLBI. Here, we explore the
  utility of Gaia data - now and in future data releases - for
  obtaining the proper motions of the more distant isolated Local Group dwarf
  galaxies in {\it Solo}, to better determine their orbital histories.

Section 2 discusses the datasets that are used in our analysis and the
identification of the dwarf galaxies for which we are able to measure
proper motions. These measurements are made in Section 3. Section 4
uses these new measurements to explore the orbital parameter space of
these galaxies (including analysis for the three dwarf galaxies for
which literature estimates of the proper motion are
available). Specifically, we examine the likelihood that these systems
are backsplash galaxies.  Section 5 discusses these findings and
summarises our results.

\section{Data and quality control}


\begin{table*}
\begin{center}
 \begin{tabular*}{\textwidth}{l|rrrrrrrr}
Galaxy & RA (degs) & Dec (degs) & $(m - M)_0$ (mags) & $M_V$ (mags) & $v_h$
                                                (km\,s$^{-1}$)&$d_{MW}$
                                                                (kpc) & $d_{M31}$
                                                                (kpc) 
   & Recent SF?\\
   \hline\\

  Eridanus 2  &  56.0879  &  -43.5333  & $  22.9  \pm  0.2  $ &  -7.2  & $  75.6  \pm  3.3  $ & 382 & 895 & N\\
  Phoenix  &  27.7762  &  -44.4447  & $  23.06  \pm  0.12  $ &  -9.9  & $  -21.2  \pm  1.0  $ & 409 & 865 & Y\\
  Leo T  &  143.7225  &  17.0514  & $  23.1  \pm  0.1  $ &  -7.6  & $  38.1  \pm  2.0  $ & 422 & 991 & Y\\
  NGC 6822  &  296.2358  &  -14.7892  & $  23.31  \pm  0.08  $ &  -15.2  & $  -54.5  \pm  1.7  $ & 452 & 898 & Y\\
Andromeda XXVIII  &  338.1717  &  31.2161  & $  24.1 ^{+ 0.5 }_{- 0.2 } $ &  -8.5  & $  -326.2  \pm  2.7  $ & 661 & 368 & N\\
  IC 1613  &  16.1992  &  2.1178  & $  24.39  \pm  0.12  $ &  -15.2  & $  -231.6  \pm  1.2  $ & 758 & 520 & Y\\
  Cetus  &  6.5458  &  -11.0444  & $  24.39  \pm  0.07  $ &  -11.3  & $  -83.9  \pm  1.2  $ & 756 & 680 & N\\
  Andromeda XXXIII  &  45.3483  &  40.9883  & $  24.44  \pm  0.18  $ &
                                                                      -10.3
                                                      &  --- & 779 & 349 & N\\
  Leo A  &  149.8604  &  30.7464  & $  24.51  \pm  0.12  $ &  -12.1  & $  24.0  \pm  1.5  $ & 803 & 1200 & Y\\
  Tucana  &  340.4567  &  -64.4194  & $  24.74  \pm  0.12  $ &  -9.5  & $  194.0  \pm  4.3  $ & 883 & 1356 & N\\
  Peg DIG  &  352.1512  &  14.7431  & $  24.82  \pm  0.07  $ &  -12.2  & $  -179.5  \pm  1.5  $ & 921 & 474 & Y\\
  WLM  &  0.4925  &  -15.4608  & $  24.85  \pm  0.08  $ &  -14.3  & $  -130.0  \pm  1.0  $ & 933 & 836 & Y\\
  Sag DIG  &  292.4958  &  -17.6808  & $  25.14  \pm  0.18  $ &  -11.5  & $  -78.5  \pm  1.0  $ & 1059 & 1357 & Y\\
  Aquarius  &  311.7158  &  -12.8481  & $  25.15  \pm  0.08  $ &  -10.7  & $  -137.7  \pm  2.1  $ & 1066 & 1172 & Y\\
  Andromeda XVIII  &  0.5604  &  45.0889  & $  25.42 ^{+ 0.07 }_{- 0.08 } $ &  -9.2  & $  -332.1  \pm  2.7  $ & 1217 & 453 & N\\
  UGC 4879  &  139.0092  &  52.84  & $  25.67  \pm  0.04  $ &  -12.5  & $  -29.2 ^{+ 1.6 }_{- 1.3} $ & 1367 & 1395 & Y\\
\end{tabular*}
\caption{All {\it Solo} dwarf galaxies in the Local Group and relevant
  parameters, including their equatorial coordinates, distance moduli, absolute magnitudes, heliocentric
  velocities and approximate distances from the Milky Way and
  M31. Parameters are taken from the
updated, online compilation from McConnachie (2012). Also indicated is whether the galaxy has
  any significant recent star formation ($\lesssim 0.1 - 1$\,Gyr; see \citealt{weisz2014}).}\label{gals}
\end{center} 
\end{table*}

{\it Solo} is a volume limited, wide field imaging survey of all nearby and
isolated dwarf galaxies, and a general introduction to the survey
is given in Paper I. Briefly, {\it Solo}
targets all known dwarf
galaxies fainter than the Magellanic Clouds which are within 3\,Mpc of the Sun and more than 300\,kpc from either M31
or the Milky Way (for a current total of 44 galaxies). Galaxies are observed with
either CFHT/Megacam (\citealt{boulade2003}) in the northern hemisphere
or Magellan/Megacam (\citealt{mcleod2015}) or
IMACS (\citealt{dressler2011}) in the southern hemisphere. Some targets are observed with
multiple instruments for calibration purposes. The total survey area
per galaxy is approximately one square degree, regardless of
telescope/instrument (for Magellan, multiple pointings are used to
cover this area, whereas only a single pointing is required for
CFHT). All dwarfs were observed in the $g-$ and $i-$bands, and some
galaxies also have $u-$band (although in what follows we do not use
the $u-$band).

Table~\ref{gals} lists the 16 {\it Solo} dwarf galaxies in the Local
Group (that is, at or within the zero-velocity survey; see
\citealt{higgs2020}, hereafter Paper II), ordered by distance
modulus. Parameters are taken from the
updated, online compilation from
\cite{mcconnachie2012}\footnote{\url{http://www.astro.uvic.ca/~alan/Nearby_Dwarf_Database.html}}. Their
distances from the center
of the Milky Way and from M31 are also given in Table~\ref{gals} assuming that the Sun is at a
distance of 8.122\,kpc (\citealt{gravity2018}) from the Galactic Center, and that M31 is at a
distance of 783\,kpc (\citealt{mcconnachie2005a}). For each of the
galaxies in Table~\ref{gals}, we have wide-field, ground-based
photometry, the majority of which was presented in Paper II. The
reader is referred to Papers I and II for details of all data
acquisition, reduction and processing.

For the three closest galaxies (Eridanus 2, Phoenix and Leo T), the
brightest red giant branch stars in these systems are luminous
enough to have reliable astrometry in Gaia DR2, and
proper motions for these three systems have previously been measured
(\citealt{fritz2018, pace2019}; \citealt[hereafter MV2020]{mcconnachie2020}). For the remaining systems,
the tip of the RGB is fainter than the Gaia DR2 detection limit. For
those galaxies with only old or intermediate-age stellar populations -
specifically  Andromeda~XXVIII, Cetus, Andromeda~XXXIII (Perseus), Tucana, and Andromeda XVIII
 - none of their
stars are bright enough to have reliable astrometry in Gaia
DR2. However, for those galaxies
which also have recent star formation, it is possible for
both blue and red supergiants to be (significantly) more luminous than
the tip of the RGB, depending on their mass and age (e.g., see
the original spectroscopic studies of supergiants in NGC 6822 and WLM
by \citealt{venn2001, venn2003}). Indeed,
this trait has allowed the determination of the proper motions of
M31 and M33 using Gaia DR2 (\citealt{vandermarel2019}), despite both
galaxies being located at approximately 800\,kpc from the Milky Way.

For the 8 galaxies in Table~\ref{gals} which have some evidence of
younger stellar populations (e.g., see Paper II, \citealt{weisz2014}) and which do not already have derived
proper motions, we cross-match our ground-based photometry to Gaia
DR2. Specifically, we only select those cataloged objects that are
classified as stellar (i.e., within $2 - \sigma$ of the locus of point
sources) in each of the $g$ and $i-$bands. In all cases, sources are
matched to within a few tenths of an arcsecond or better. For the Gaia
parameters, we only consider stars with full 5 parameter astrometric
solutions, and high quality astrometry as defined via the renormalised
unit weighted error ({\tt ruwe}; see \citealt{lindegren2018} and
discussion in the Gaia DR2 Documentation Release 1.2). We adopt {\tt
  ruwe} $ < 1.4$. We also only consider stars whose parallaxes are
consistent with them being located at a similar distance to the
galaxy. That is, the $3-\sigma$ parallax range measured by Gaia DR2
must overlap the $3-\sigma$ parallax range implied by the distance
modulus of the dwarf as given in Table~\ref{gals}. We correct for the global zero-point of the parallax in Gaia DR2 of -0.029mas (\citealt{lindegren2018}).

\begin{figure*}
    \includegraphics[width=0.4\textwidth]{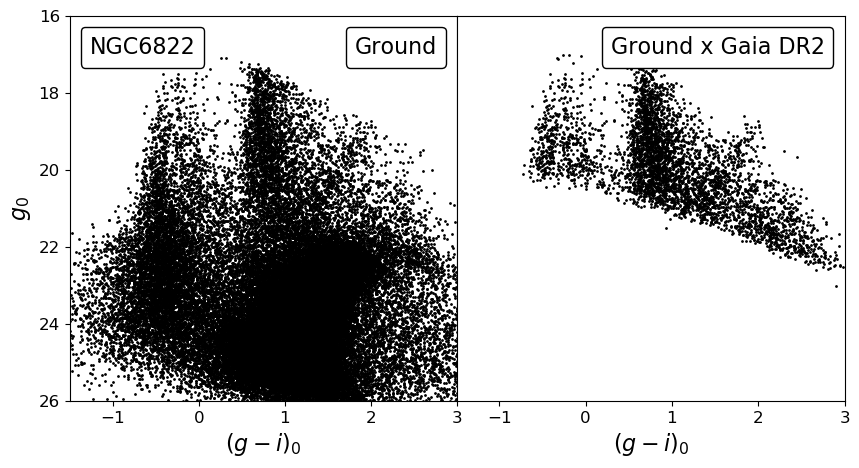}
    \includegraphics[width=0.4\textwidth]{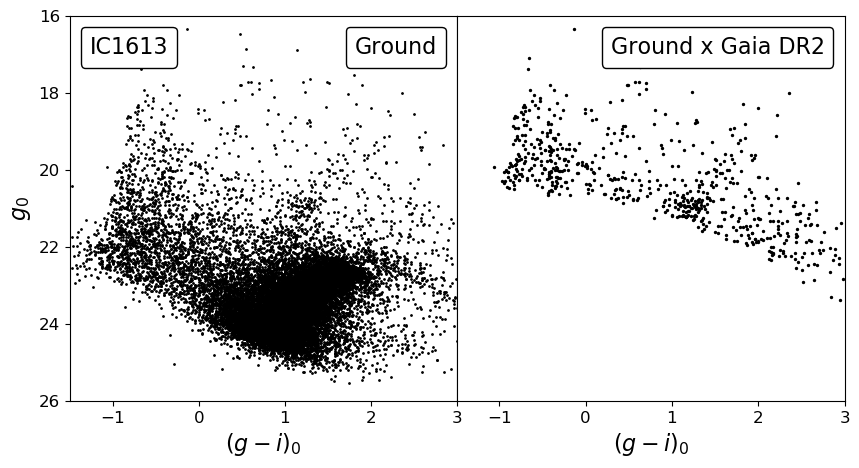}
    \includegraphics[width=0.4\textwidth]{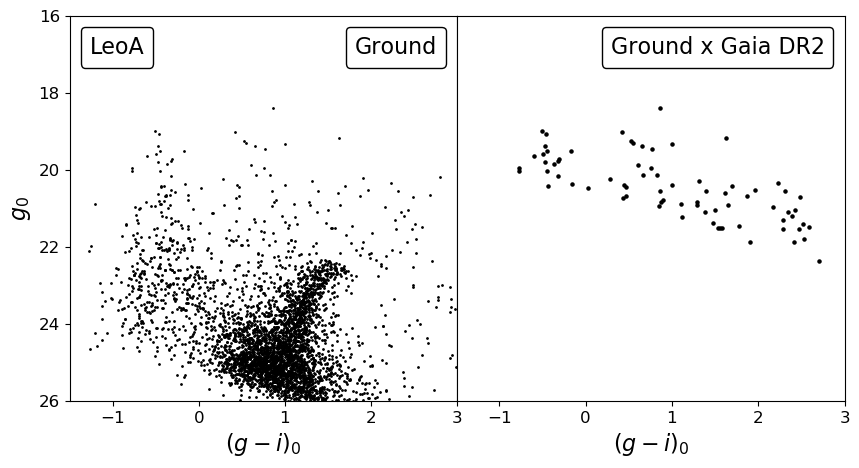}
    \includegraphics[width=0.4\textwidth]{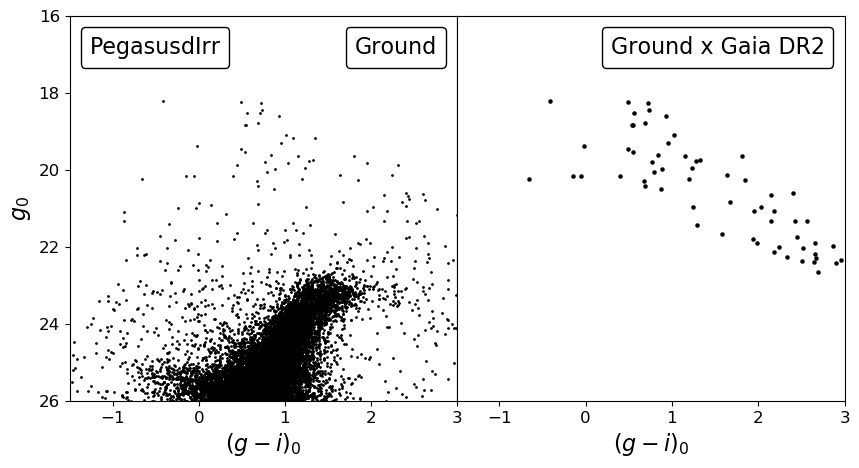}
    \includegraphics[width=0.4\textwidth]{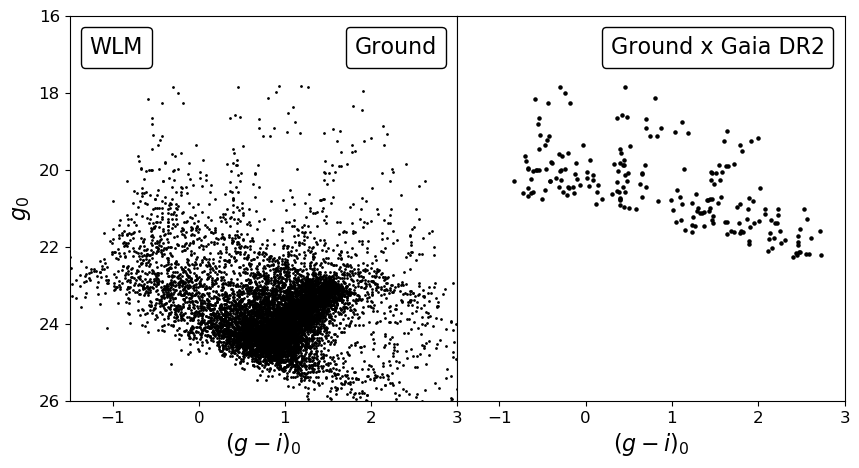}
    \includegraphics[width=0.4\textwidth]{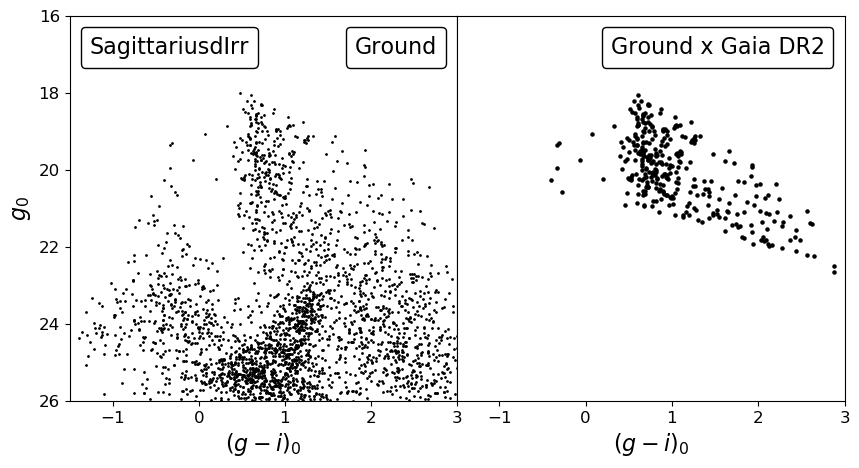}
    \includegraphics[width=0.4\textwidth]{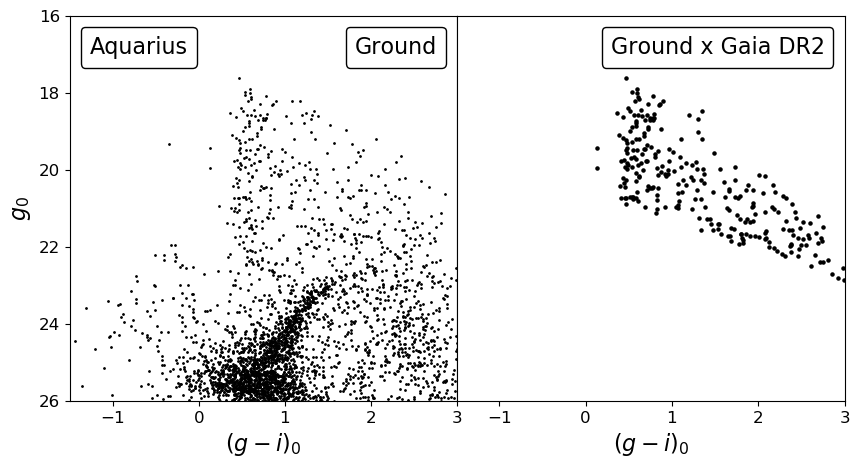}
    \includegraphics[width=0.4\textwidth]{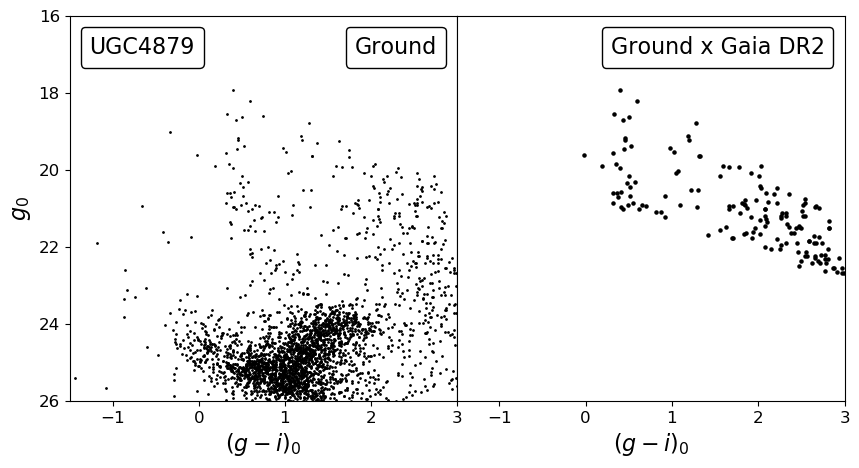}
    \caption{Extinction corrected, ground-based colour magnitude
      diagrams (CMDs) for each of the galaxies listed in
      Table~\ref{gals} that do not already have Gaia DR2 systemic proper motions, but
      which have some evidence of younger stellar populations. The
      left panel of each pair of CMDs shows the ground-based CMDs for
      the inner regions of each field centered on each galaxy. The right
      panel of each pair shows those sources that are also in Gaia
      DR2, and which pass the quality cuts described in the
      text.}\label{cmds}
  \end{figure*}

The ground-based photometry is presented in Figure~\ref{cmds}. Here, the left panel of each pair
of plots shows the ground-based CMD for the inner regions of each
field centered on each galaxy. The right panel of each pair shows those sources that
are also in Gaia DR2, and which pass the cuts described above. All
photometry is extinction-corrected with the reddening maps from
\cite{schlegel1998} using the  python interface {\tt
  dustmaps} (\citealt{green2018}). We use the extinction coefficients
derived using the Padova isochrone (\citealt{girardi2002}) web-tool\footnote{http://stev.oapd.inaf.it/},  $A_g = 1.191\times A_V$
and $A_i = 0.854 \times A_V$,
where $A_V = 2.742 \times E(B - V)$ (\citealt{schlafly2011}).

In contrast to MV2020, we do not apply
any cuts to the Gaia DR2 data based on the quality of the Gaia
photometry, since we rely instead on the ground-based data for our
photometric selection. This is an important point: experimentation
with the Gaia DR2 photometry shows that the formal quality of the Gaia
photometry for many of the sources being examined is quite poor (as
quantified via the {\tt fluxexcess} parameter; see
\citealt{lindegren2018}). We expect this is to do with the relatively
high degree of crowding in these distant dwarf galaxies (see also
discussion of this point in \citealt{vandermarel2019} as it relates to
the centers of M31 and M33). Of course, the
ground-based data can also be crowded in the very central regions of
the dwarfs, especially at faint magnitudes. However, the generally
good or excellent image quality of our ground-based data, in multiple bands,
combined with the relative brightness of the sources under
consideration (with respect to the limiting magnitude of the
ground-based data) means that the photometric quality of Gaia DR2 is not
a limiting factor for this analysis, so long as the astrometric
quality of the data is sufficient (as parameterized by {\tt ruwe}).

For the 8 galaxies in Figure~\ref{cmds}, blue supergiants are easily
visible as a broad vertical plume of stars generally blueward of
$(g - i)_0 = 0$ in the left panels (although UGC 4879 in particular
does not have many such stars; see also \citealt{jacobs2011}). In all cases, they extend to
luminosities brighter than the tip of the RGB. However, not all
galaxies have many of these stars bright enough to have robust
astrometry in Gaia DR2. Specifically, Peg DIG, Sag DIG, Aquarius
and UGC 4879 do not have more than five or six candidate blue
supergiants (if any) that remain in the right panels of
Figure~\ref{cmds}. In addition to blue supergiants, red supergiants
are clearly visible in NGC 6822, IC 1613 and WLM, and form a diagonal
sequence at a colour of around $(g - i)_0 \simeq 1$ extending in the
direction of the top-right corner.  Equivalent sequences are not
clearly visible in any of the other galaxies. Leo A is a marginal case
with respect to the other systems since blue supergiants are present
in the Gaia DR2 data but there is no clear red supergiant
sequence.

The right panels of Figure~\ref{cmds} suggest that systemic proper
motions should be able to be derived using Gaia DR2 data for NGC 6822,
IC 1613, Leo A and WLM. We make these measurements in Section~\ref{pms}.

\section{Measurement of proper motions}\label{pms}

\subsection{Method}

\begin{figure*}
    \includegraphics[width=\columnwidth]{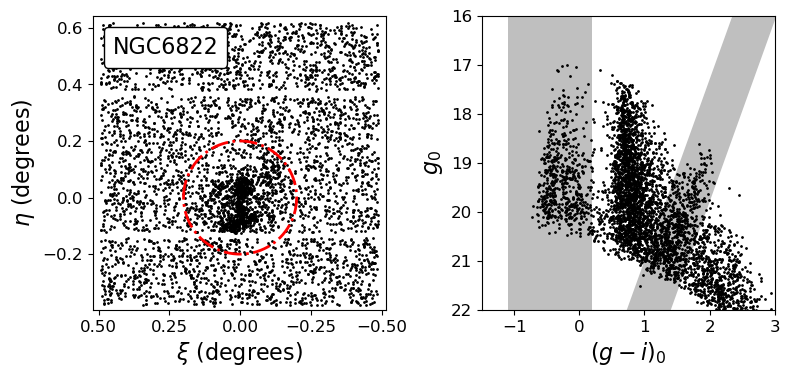}
    \includegraphics[width=\columnwidth]{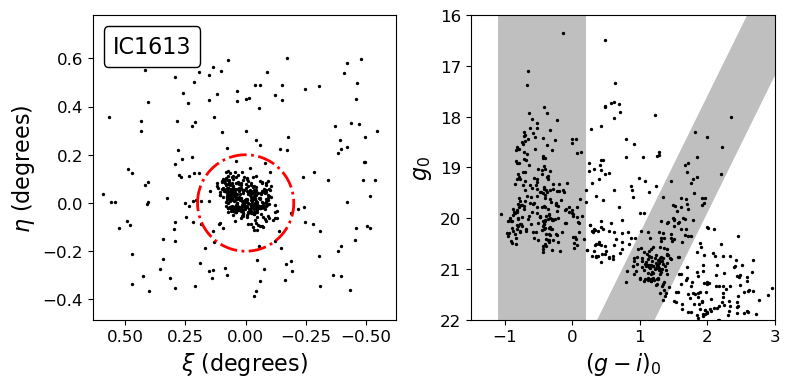}
    \includegraphics[width=\columnwidth]{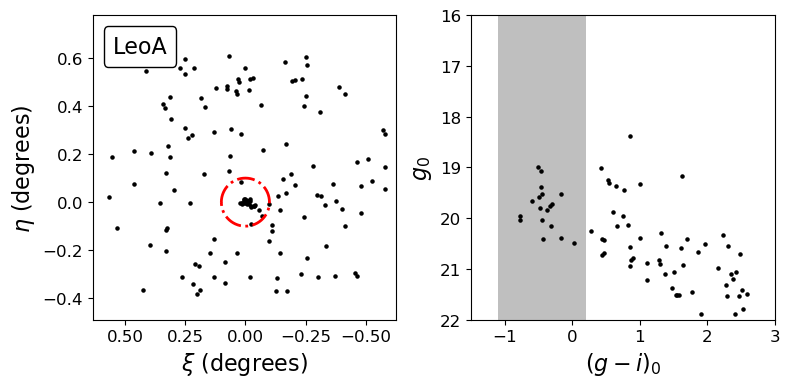}
    \includegraphics[width=\columnwidth]{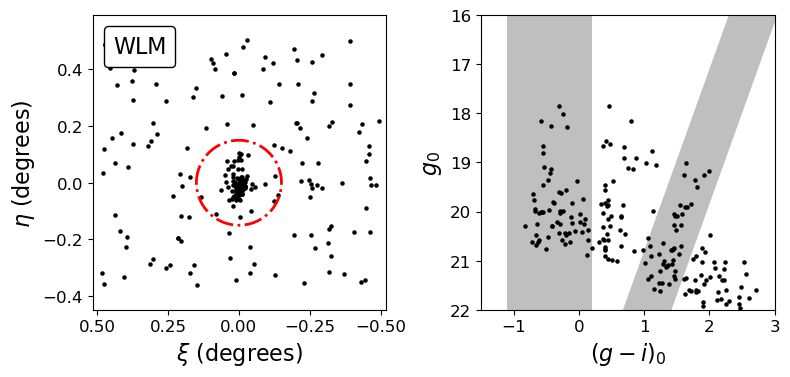}
    \caption{The spatial and colour-magnitude distributions of Gaia DR2
      sources in each field surrounding NGC 6822, IC 1613, Leo A and WLM. The
      left panels show the distribution in the tangent plane of all
      Gaia DR2 sources that lie within the shaded regions (blue and
      red supergiant regions) of the CMD
      in the right panels. The right panel shows the CMD of all Gaia
      DR2 sources that lie within the dashed red circles in the left
      panels.}\label{cmdspatial}
\end{figure*}

We adopt a similar methodology to MV2020, and we refer the reader to
Section 3 of that paper for a full description of the method. In
brief, we seek to maximize the likelihood of the data, where the total
likelihood is given by

\begin{equation}
\mathcal{L} = f\textsubscript{dwarf}\mathcal{L}\textsubscript{dwarf} + (1 - f\textsubscript{dwarf})
\mathcal{L}_{MW} \label{maxl}
\end{equation}

\noindent where $\mathcal{L}\textsubscript{dwarf}$ and $\mathcal{L}_{MW}$ are the likelihoods
for the dwarf galaxy and MW foreground, respectively, and
$f\textsubscript{dwarf}$ is the fraction of stars in the satellite. $\mathcal{L}\textsubscript{dwarf}$ can be broken down as

\begin{equation}
\mathcal{L}\textsubscript{dwarf} = \mathcal{L}_{s}\mathcal{L}_{CM}\mathcal{L}_{PM} \label{breakdown}
\end{equation}

\noindent where $\mathcal{L}_s, \mathcal{L}_{CM}$ and
$\mathcal{L}_{PM}$ are the likelihoods from the spatial information,
colour-magnitude information, and proper motion information,
respectively. $\mathcal{L}_{MW}$ can similarly be broken down into the
product of its three constituent likelihoods.

We therefore need to define likelihood functions for the spatial and
colour-magnitude distribution of sources in the dwarf. We also need to
define spatial, colour-magnitude and proper motion likelihood functions
for the Milky Way foreground population. Using these, we find the
systemic proper motion of the dwarf that maximises the probability of
the data given these models, with the adoption of suitable priors on
the unknown parameters $(\mu_\alpha\cos\delta, \mu_\delta)\textsubscript{dwarf}$ and $f\textsubscript{dwarf}$.

Figure~\ref{cmdspatial} shows the spatial and colour-magnitude
distributions of Gaia DR2 sources in each field surrounding NGC 6822,
IC 1613, Leo A and WLM. Specifically, the left panels show the distribution
in the tangent plane of all Gaia DR2 sources that lie within the
shaded regions of the CMD in the right panels. The right panel shows
the CMD of all Gaia DR2 sources that lie within the dashed red
circles in the left panels. The grey regions are defined empirically to be simple
polygons that enclose the blue and red supergiants in each galaxy. For
Leo A, we only consider blue supergiants due to the absence of any
clear red supergiant sequence. For the blue
supergiants, the red edge of the selection box is placed bluer than
the edge of the Milky Way foreground (that is delineated by the main
sequence turn-off of stars around $(g - i)_0 \sim 0.4$ at a range of distances in the halo of the Milky Way).

For the spatial, colour-magnitude and proper motion likelihood
functions for the Milky Way foreground population, the corresponding
likelihood functions are defined in exactly the same way as in
MV2020. That is, for the spatial likelihood function, we consider the
contamination to be uniform across the field. For the colour-magnitude
and proper motion likelihood functions, we use the relevant
distributions of stars that are outside of the red dashed circles in
Figure~\ref{cmdspatial}. These are converted into density maps and
normalized, as described in MV2020.

For the spatial likelihood function of the satellite, MV2020 used
2D-lookup maps based on the parameterized structure of the galaxies as
reported in the literature. However, in general these parameters are
derived from the distribution of old and intermediate age stellar
populations (usually upper main sequence and/or red giant branch
stars; e.g. \citealt{munoz2018}), and these can have a very different
spatial distribution to the young stars that are the focus
here (for example, note the irregular distribution of young stars in
NGC 6822). In the absence of good parameterizations
of the spatial distribution of young stars, we instead build a
likelihood function based on the projected density distribution of
candidate blue supergiants. A density map (0.1 arcmin
pixels) is produced using stars that lie within the grey regions corresponding to the
blue supergiant locus for each galaxy, and that are within the dashed red
circles in Figure~\ref{cmdspatial}. These density distributions are
then convolved with a Gaussian with a standard deviation of 0.5
arcmins, and normalized appropriately. We rely on blue
supergiants to define the spatial likelihood function because they
can be identified with almost no contamination from foreground
populations. The red supergiants in NGC6822, IC1613 and WLM are
assumed to be spatially distributed in much the same way as the blue
supergiants, which seems reasonable

For the CMD likelihood function of the satellite, MV2020 used
2D-lookup maps based on the expected location of upper main sequence,
sub-giant and red giant branch stars as given by isochrones with
appropriate ages and metallicities. In comparison, for blue and red
supergiants, the expected CMD distribution is not as sharply defined
and they are more difficult to model without reasonably detailed
recent star formation histories (e.g., see the beautiful
reconstruction of the star formation history of Sextans~A using blue
and red supergiants by \citealt{dohmpalmer2002}). Instead, we instead use a simple,
empirical model of the expected CMDs, which is that the relevant stars
are expected to lie somewhere in the grey regions of the CMDs in
Figure~\ref{cmdspatial}. A uniform likelihood is assumed over
these regions.

In MV2020, two different sets of priors were imposed on the systemic
proper motion of the dwarf. The first was that the tangential velocity
could not be unrealistically high compared to the Milky Way; the
second favored systemic motions where stars that appeared to be radial
velocity members were more likely to be assigned membership. In the
current analysis, there are no radial velocities to consider, and we
do not have any expectation on the tangential velocity of the galaxy
with respect to the Milky Way. As such, the prior requires only that
the systemic proper motion in each direction is less than
$10\,$mas/yr, with a uniform likelihood over this range. We
adopt a uniform prior for $f\textsubscript{dwarf}$ in the interval
$0 \le f\textsubscript{dwarf} \le 1$, the same as MV2020.

\subsection{Results}\label{results}

\begin{table*} 
\begin{center}
\begin{tabular*}{\textwidth}{l|rrrrrrrr}
Galaxy &$\mu_\alpha\cos\delta$ (mas/yr)&$\mu_\delta$ (mas/yr)&$f\textsubscript{dwarf}$ &  $v_\alpha\cos\delta$ (km\,s$^{-1}$) & $v_\delta$ (km\,s$^{-1}$)&$v_t$ (km\,s$^{-1}$)&$v_r$ (km\,s$^{-1}$) & Source\\
\hline 
Eridanus 2 &   $ 0.35 ^{+0.21}_{-0.20}$   &   $ -0.08 \pm 0.25 $ & $
                                                                  0.0027
                                                                  ^{+0.0008}_{-0.0007}$&
                                                                                         $472
                                                                                         ^{+378}_{-360}$
                                                                                                                 &  $ -40 \pm 451 $ & 473 & -73 &
  MV2020\\ 
Phoenix &   $ 0.08 \pm 0.15 $   &   $ -0.08 \pm 0.18 $  & $ 0.009 \pm
                                                           0.002 $ & $ -2 \pm 291 $ &  $ 3 \pm 349 $ & 3 & -114 & MV2020 \\ 
Leo T &   $ 0.10^{+0.67}_{-0.69} $   &   $ 0.01 ^{+0.57}_{-0.56} $  & $ 0.002 \pm
                                                        0.001 $ &$ 227 ^{+1324}_{-1364} $
                                         &  $ 237 ^{+1126}_{-1107} $ & 328& -63 & MV2020\\ 
  NGC 6822 & $-0.02\pm 0.02$ & $ 0.00 \pm 0.02$ & $0.73 \pm
                                                         0.01$& $5 \pm
                                                                                                                 44$
                                                                                                                 &
                                                                                                                                            $212
                                                                                                                                            \pm
                                                                                                                                            44
                                                                                                                                            $&
                                                                                                                                               212
                                                                                                                                               &
                                                                                                                                                 51 &
                                                                                                                   This paper\\
  IC 1613 & $0.08 ^{+0.06}_{-0.07} $ & $-0.04 \pm 0.03 $ & $0.96 \pm
                                                          0.01$  &
                                                                   $141^{+215}_{-251}$&
                                                                                        $32\pm107$&
                                                                                                    145
                                                                                                                                                                 & -150 &
                                                                                                                   This paper\\
  Leo A & $ 0.49 \pm 0.32$ & $ -0.46 \pm 0.28 $ & $0.96_{-0.05}^{+0.03}$
  & $1900 \pm 1211$ & $-1507 \pm 1059$ & 2425 & -20 & This paper\\
  WLM & $ 0.11 \pm 0.08$ & $ -0.10 \pm 0.05 $ & $0.93_{-0.03}^{+0.02}$
  & $356 \pm 346$& $-231 \pm 216$ & 425& -70 &  This paper\\
\end{tabular*}
\caption{Median, 16th and 84th percentiles of the probability density
  functions for the three unknown parameters for all {\it Solo} dwarf
  galaxies in the Local Group for which derivation of a systemic
  proper motion is currently possible (Table 4 of MV2020 presents values for
  Eridanus 2, Leo T and Phoenix; the current paper derives values for
  NGC 6822, IC 1613 and WLM using a modification of the same method).  The corresponding tangential velocity components in a Galactocentric frame of reference 
 are listed, as well as the overall tangential velocity (v$_t$). 
The implied Galactocentric radial velocities are listed in the last
column for comparison, and are converted from the heliocentric radial
velocities listed in Table~\ref{gals} (see text for details).}
\label{pmresults}
\end{center} 
\end{table*}

\begin{figure*}
    \includegraphics[width=0.75\textwidth]{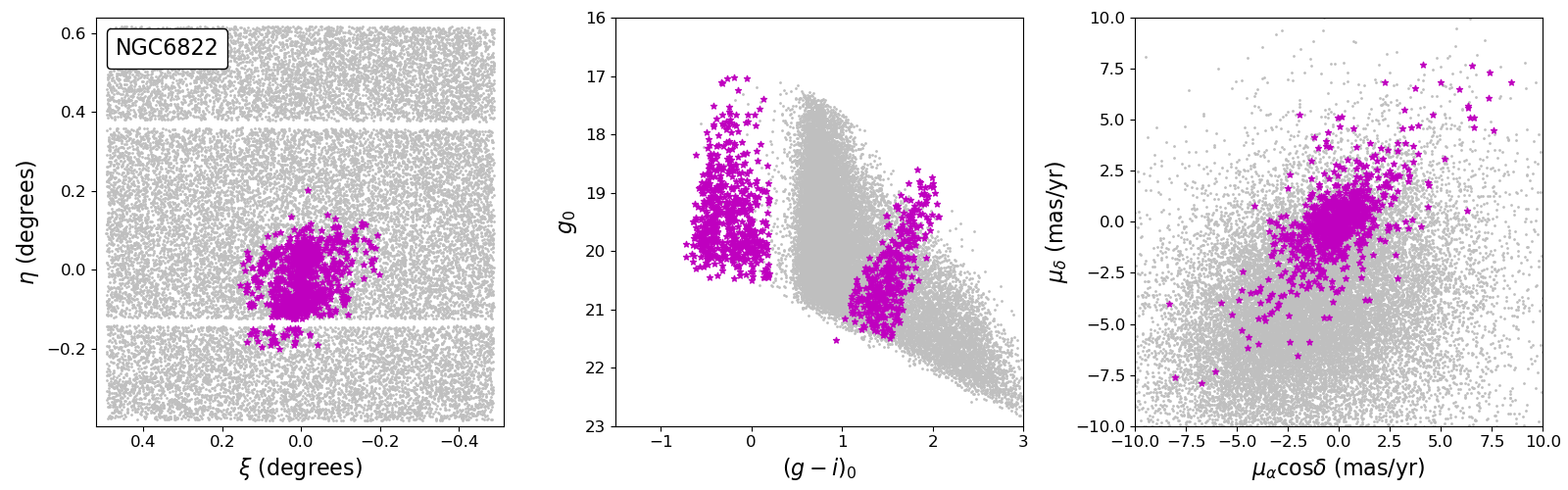}
    \includegraphics[width=0.75\textwidth]{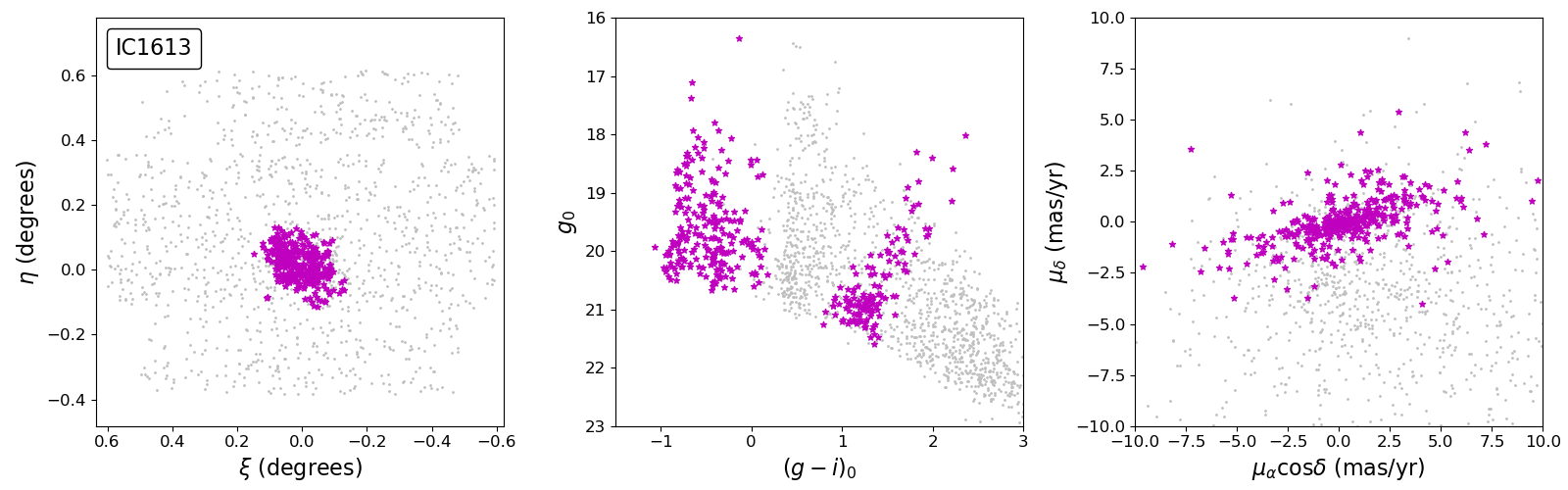}
    \includegraphics[width=0.75\textwidth]{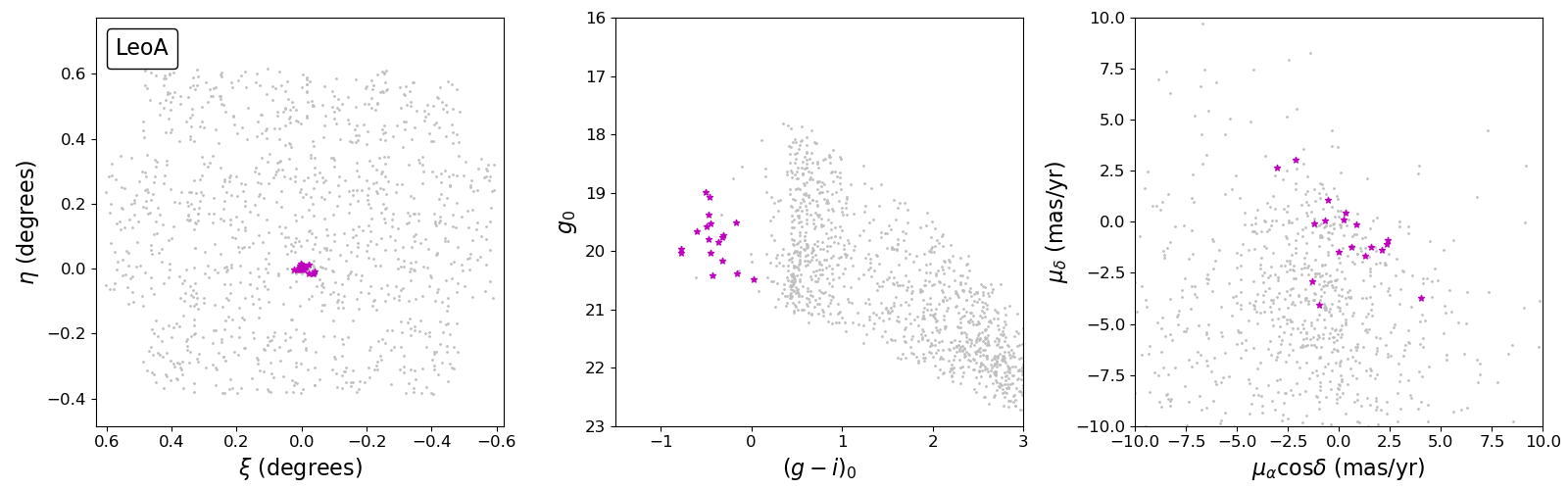}
    \includegraphics[width=0.75\textwidth]{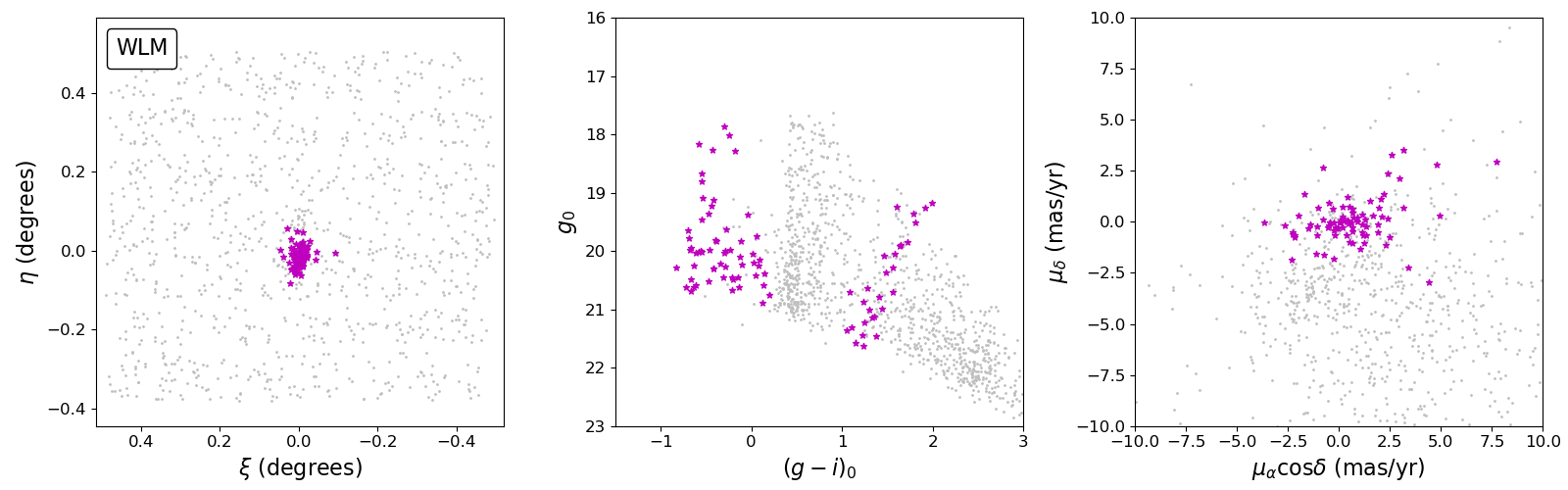}
    \caption{The spatial (left panels), colour-magnitude (middle panels), and proper
motion (right panels) distributions for member stars (magenta stars)
in NGC 6822, IC 1613, Leo A and
WLM. Grey points show all Gaia DR2 sources that pass our quality-control criteria
described in the text.}\label{members}
\end{figure*}

For NGC 6822, IC 1613, Leo A and WLM, we construct the spatial, colour-magnitude
and proper motion likelihood maps as described in the previous section
and explore parameter space using {\tt emcee}
(\citealt{foremanmackey2013, foremanmackey2019}) for those values of $f\textsubscript{dwarf}$ 
and $\left(\mu_\alpha\cos\delta, \mu_\delta\right)\textsubscript{dwarf}$ that maximise the probability of the data.

Table~\ref{pmresults} lists the  resulting median values and 16th and 84th
percentiles for the three unknown parameters for each of our four
galaxies. We also list the proper motions for Eridanus 2, Phoenix and
Leo T, as derived in MV2020\footnote{Note that the relative foreground
  for these three galaxies is much higher than that for the dwarf
  irregulars studied here due to differences in the
  tracer population that is used (RGB stars versus supergiants). This
  results in very different values for $f\textsubscript{dwarf}$ between the two studies.}. We also
list the corresponding Galactocentric tangential velocity components
assuming $(R_{\odot}, V_c) = (8.122$\,kpc, 229\,km\,s$^{-1})$
and 
$(U_\odot, V_\odot, W_\odot) = (11.1, 12.24, 7.25)$\,km\,s$^{-1}$
(\citealt{gravity2018, schonrich2010}), for the distance moduli given
in Table~\ref{gals}.

All error bars in Table~\ref{pmresults} describe random errors only, and systematic
uncertainties in the proper motions are not included. \cite{helmi2018b}
analyse the systematic uncertainties in the proper motions of stars in
the Sagitarrius dwarf spheroidal using Gaia DR2, and conclude that on spatial scales
of around 0.2 degrees (the approximate spatial scale of the
supergiant distributions in NGC 6822 and IC 1613), the systematic uncertainty is  
$\sim 0.035$\,mas/year. \cite{lindegren2018} analyse the spatial
covariance of the proper motions of quasars, and conclude that on
scales less than 0.125 degrees (the approximate spatial scales of all
the other galaxies in Table~\ref{pmresults}, excluding NGC 6822 and IC
1613), the systematic uncertainty is
$\sim 0.066$\,mas/yr. It
is worth emphasising the scale of these uncertainties for the galaxies
under consideration. At the distances of NGC 6822 (462\,kpc) and WLM
(933\,kpc), a systematic uncertainty of [0.035/0.066] mas/year corresponds to
an uncertainty in the tangential velocity of these dwarfs of [73/138]\,km\,s$^{-1}$
and [148/280]\,km\,s$^{-1}$, respectively, in each direction. As a reference
value, a simple estimate  of the expected velocity
dispersion of the Local Group gives $\sigma_{LG} \simeq 161 \left(\frac{M_{LG}}{ 3 \times 10^{12}
    M_\odot}\right)^{\frac{1}{2}}\left(\frac{500\,kpc}{
    R_{LG}}\right)^{\frac{1}{2}}$\,km\,s$^{-1}$.

Figure~\ref{members} shows the spatial, colour-magnitude, and proper
motion distributions for member stars in NGC 6822, IC 1613, Leo A and
WLM (magenta stars). Here, we use the fact that the probability of any star being a
member of the dwarf is given by

\begin{equation}
 P\textsubscript{dwarf} =  \frac{f\textsubscript{dwarf}\mathcal{L}\textsubscript{dwarf}}{f\textsubscript{dwarf}\mathcal{L}\textsubscript{dwarf} +
   (1 - f\textsubscript{dwarf})  \mathcal{L}_{MW}}~.
\end{equation}

\noindent We define members as those stars with $ P\textsubscript{dwarf} \ge
0.5$. Inspection of Figure~\ref{members} shows that the probable
members of these galaxies appear to cluster appropriately in all three
parameter spaces, although the very few stars in Leo~A mean that there
is not as obvious a cluster of points in proper motion space as for
the other three galaxies.

\section{Orbital histories}

Here, we consider the orbits of our target dwarf galaxies within the Local Group to better understand
what constraints the newly derived systemic proper motions place on their orbital
histories. In particular, we examine the
likelihood  that these dwarfs have ever passed within 300\,kpc of either M31
or the Milky Way.

\subsection{The Local Group potential}

We explore the orbits of each of the dwarf galaxies backwards in time within the Local
Group. All our orbital calculations use {\tt gyrfalcON}
(\citealt{dehnen2000}). Our intent is to
determine only if the dwarf has passed within 300\,kpc of either the
Milky Way or M31. Each of the
dwarf galaxies starts well outside the virial radius of either the
Milky Way or M31: $R_{MW, vir} \sim 287$\,kpc assuming an NFW profile
with a total mass of $M_{MW, vir} \simeq 1.3 \times 10^{12}M_\odot$
(\citealt{posti2019}), and this should be compared to the present-day distances of the closest dwarfs (for the MW -- Eridanus 2, $d_{MW} \simeq
382$\,kpc; for M31 -- IC 1613, $d_{MW} \simeq
520$\,kpc). As such, we model each large galaxy as a point mass. {\tt gyrfalcON} does not allow for the use of
tracer particles, so we instead set the mass of the dwarf to be tiny
in comparison to either of the two large galaxies.

We adopt the distance modulus of M31 to be $d_{mod} = 24.47 \pm 0.07$
(\citealt{mcconnachie2005a}) and its heliocentric radial velocity to be
$v_h = -301 \pm 1$\,km\,s$^{-1}$ (\citealt{vandenbergh2000}). We adopt
the systemic proper motion of M31 derived using Hubble Space Telescope
observations (\citealt{sohn2012}) as given in \cite{vandermarel2012a},
namely $\mu_\alpha\cos\delta = 0.045 \pm 0.013$ mas/year, $\mu_\delta =
-0.032 \pm 0.012$ mas/year. For the dwarfs, we use
the distance moduli and heliocentric radial velocities as given in
Table~\ref{gals}, and the systemic proper motions as given in
Table~\ref{pmresults}. We add in quadrature a systemic proper motion uncertainty to
the random uncertainties given in this table. For NGC 6822 and IC 1613, we use
$\sigma_{sys} = 0.035$ mas/year (\citealt{helmi2018b}) given the relatively large angular sizes of
these system.  For the remaining galaxies, we use $\sigma_{sys} =
0.066$ mas/year (\citealt{lindegren2018}). We transform all
velocities and proper motions to a Galactocentric frame using the
parameters given in Section~\ref{results}.

There is still considerable debate over the total masses of the Milky Way
and M31. The former offers more opportunity for study, and
\cite{eadie2019} note that there is reasonable agreement in the
literature regarding the mass at small ($\lesssim 40$\,kpc) radius. More
difficult to constrain is the mass at large radius, where there are
fewer tracers and generally poorer data, and many estimates rely on
extrapolation of results obtained at smaller radius, given some model assumptions.
\cite{blandhawthorn2016} provide a good review of the relevant
literature, and calculate the average of a range of
studies 
based on analysis of the kinematics of the stellar halo. They find that
$M_{MW} \simeq (1.3 \pm 0.3) \times 10^{12}\,M_\odot$. This is an 
identical value to that obtained by \cite{posti2019}, on extrapolation
of the mass that they measure within 20\,kpc,  based upon Gaia DR2
kinematics of globular clusters. A different analysis of globular
cluster data - including but not limited to Gaia DR2 - by
\cite{eadie2019} favors a slightly lighter mass, $M_{MW} \simeq
(0.7^{+0.4}_{-0.19}) \times 10^{12}\,M_\odot$ (where the uncertainties
describe the 95\% credible intervals). \cite{cautun2020} estimate
$M_{MW} = (1.08_{-0.14}^{+0.20}) \times 10^{12}\,M_\odot$ from
analysis of the Gaia DR2 rotation curve and other data. 

For M31, its integrated
properties generally  suggest that it is more massive than the MW
(e.g, \citealt{hammer2007}). \cite{watkins2010} analysed the satellite kinematics and
determined that $M_{M31} = (1.4 \pm 0.4) \times
10^{12}\,M_\odot$. \cite{fardal2013} examined the dynamics of the
Giant Stellar Stream (\citealt{ibata2005}) and concluded that $M_{M31} \simeq (1.9 \pm 0.4) \times
10^{12}\,M_\odot$. \cite{patel2017b} combine astrometric studies with
cosmological simulations to estimate that $M_{M31} = (1.37^{+1.39}_{-0.75}) \times
10^{12}\,M_\odot$ (where they also estimate that $M_{MW} =
(1.02^{+0.77}_{-0.55}) \times 10^{12} M_\odot$). A direct measurement
of the mass ratio between M31 and the MW was made by
\cite{penarrubia2014}, who found that $f = M_{MW}/M_{M31} =
0.54^{+0.23}_{-0.17}$ by analysis of the effect of the Local Group
mass on the Hubble expansion of  its nearest neighbours.

\subsection{Exploring parameter space}

Here, we run a suite of simulations for each dwarf in which we run
their orbits back in time, varying
their two tangential velocity components across a grid, to determine
which combinations yield interactions with the Milky Way and/or
M31. We then compare this parameter space to the actual measured proper
motions of the dwarfs.

We vary the Galactocentric tangential velocity of each dwarf in the range
$-1000$\,km\,s$^{-1}$ $\le v_\alpha\cos\delta, v_\delta \le 1000$\,km\,s$^{-1}$, searching the
full grid of values in 5\,km\,s$^{-1}$ increments. We note that the velocity
dispersion of the Local Group is expected to be of order $\sigma_{LG} \simeq 161 \left(\frac{M_{LG}}{ 3 \times 10^{12}
    M_\odot}\right)^{\frac{1}{2}}\left(\frac{500\,kpc}{
    R_{LG}}\right)^{\frac{1}{2}}$\,km\,s$^{-1}$, and so this grid
search extends out to velocities that would certainly make these
galaxies unbound to the Local Group. All other positional and
dynamical parameters of the dwarf and M31 are fixed to their preferred
values as described above. For the masses of the Milky Way and M31, we
use three different combinations:

\begin{enumerate}
\item LG1: $M_{MW} = 1.3 \times 10^{12}M_\odot$, $f = 1.0$; 
\item LG2: $M_{MW} = 1.3 \times 10^{12}M_\odot$, $f = 0.5$; 
\item LG3: $M_{MW} = 1.6 \times 10^{12}M_\odot$, $f = 1.0$.
\end{enumerate}

Each of these LG realisations uses a mass for the MW that is either
somewhat ``average'' or to the upper end of the recent estimates
(see the discussion above). We purposefully explore orbits using these
large masses because, in what follows, the most conservative estimates on the chance
of an interaction between a dwarf and the MW/M31 will be obtained
under the assumption of larger masses (if an interaction is not going to occur
assuming a larger mass estimate, then it will definitely not happen
for a lighter mass, all other things being equal).

The point mass approximation for M31 and the MW is not sufficient to
determine exact pericenters for the dwarf orbits. However, we only
seek to determine if the dwarf passes within 300\,kpc of either the
Milky Way or M31 (the approximate virial radius of these
galaxies). Prior to reaching this distance from the Milky Way or M31,
the dwarf will have been affected by the gravitational influence of
essentially the total mass of the galaxy, thus the point mass
approximation is sufficient up to this point. We do not seek to make
any statements about what happens to the orbit after this threshold is
reached.

\begin{figure*}
    \includegraphics[width=0.3\textwidth]{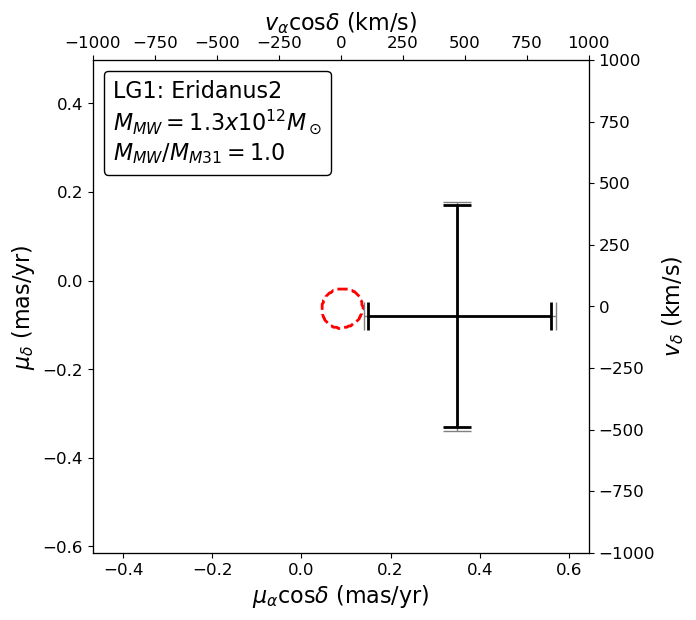}
    \includegraphics[width=0.3\textwidth]{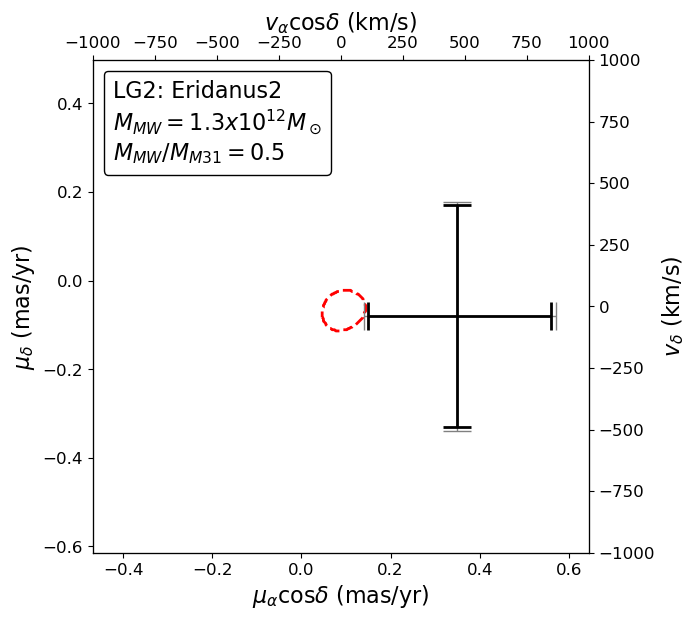}
    \includegraphics[width=0.3\textwidth]{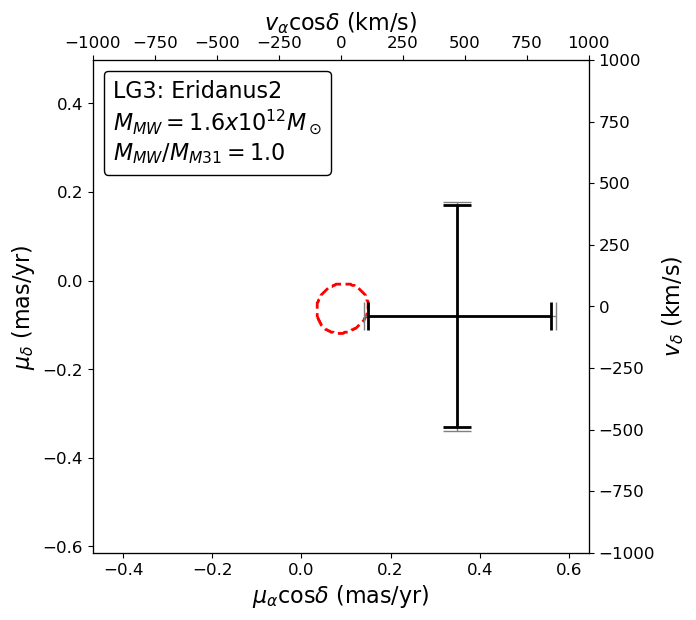}
    \includegraphics[width=0.3\textwidth]{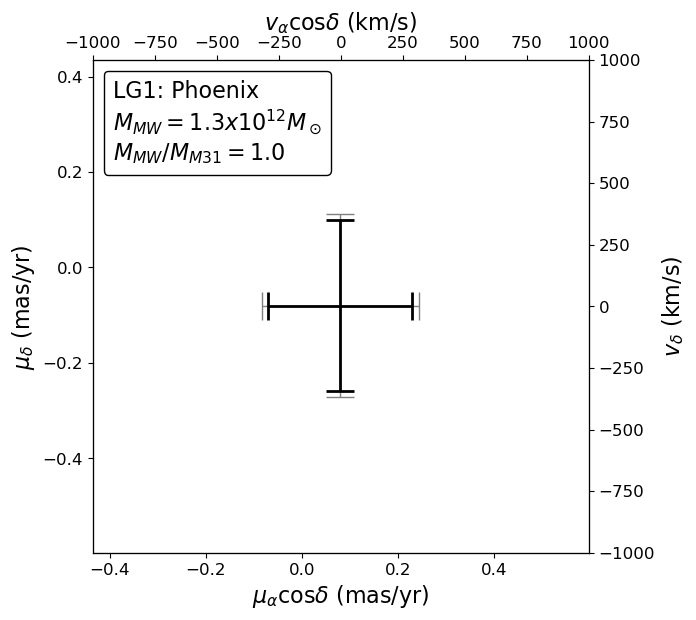}
    \includegraphics[width=0.3\textwidth]{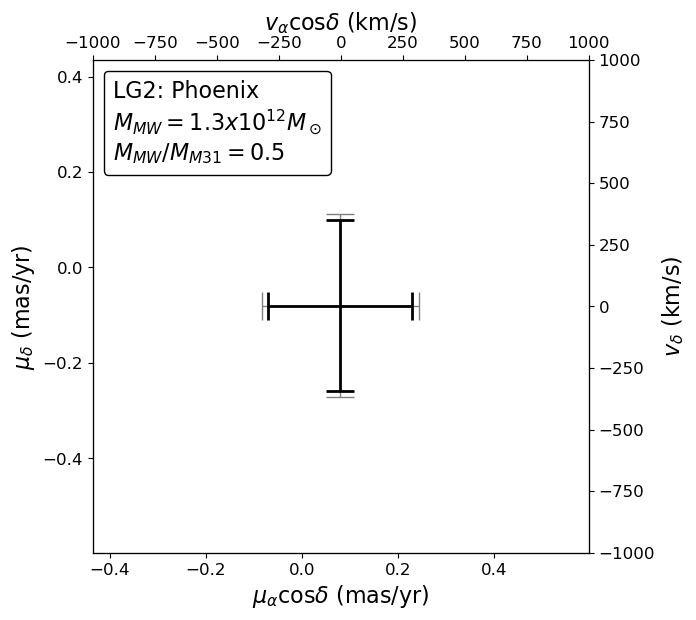}
    \includegraphics[width=0.3\textwidth]{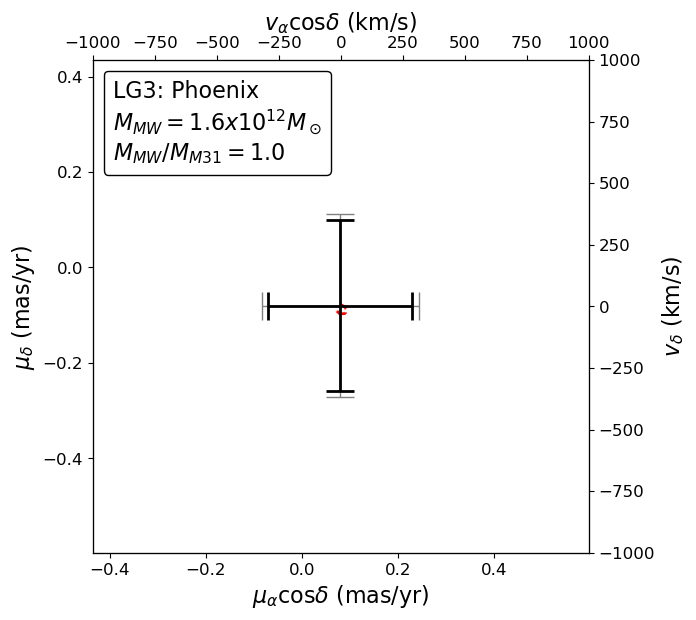}
    \includegraphics[width=0.3\textwidth]{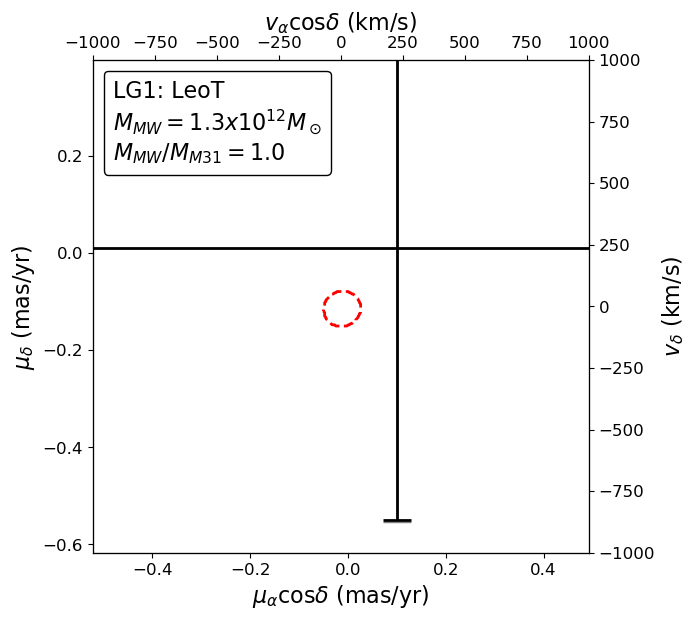}
    \includegraphics[width=0.3\textwidth]{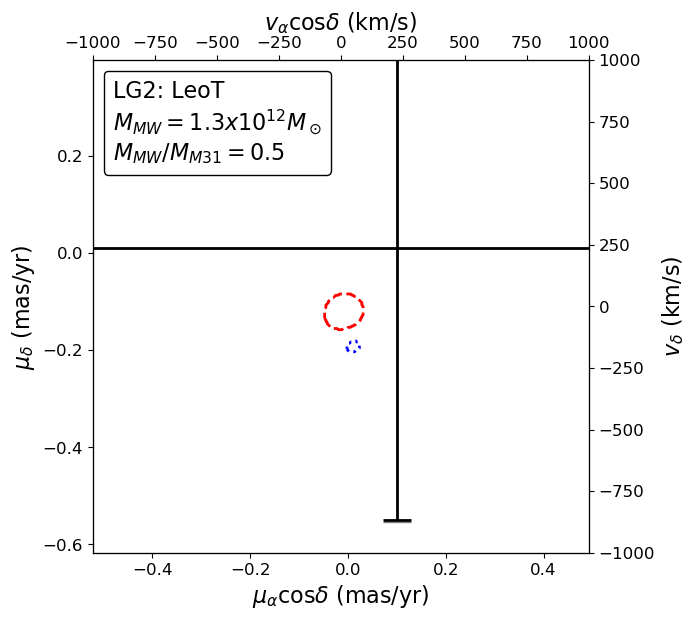}
    \includegraphics[width=0.3\textwidth]{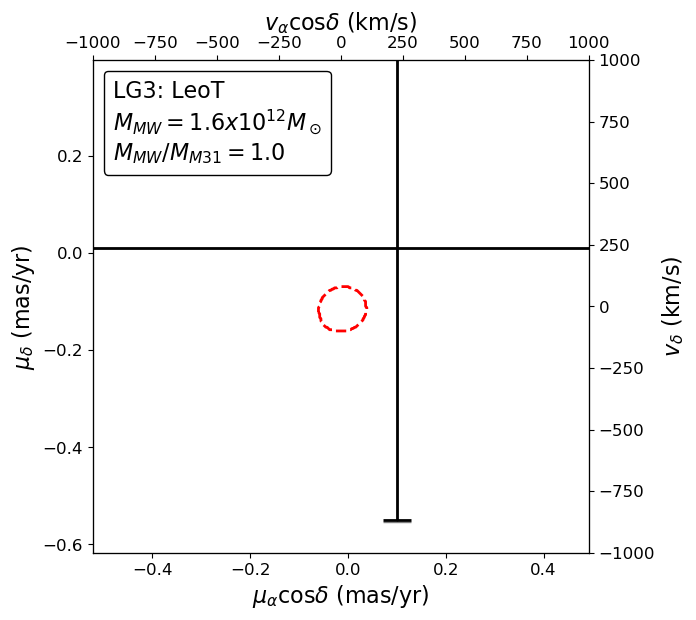}
    \includegraphics[width=0.3\textwidth]{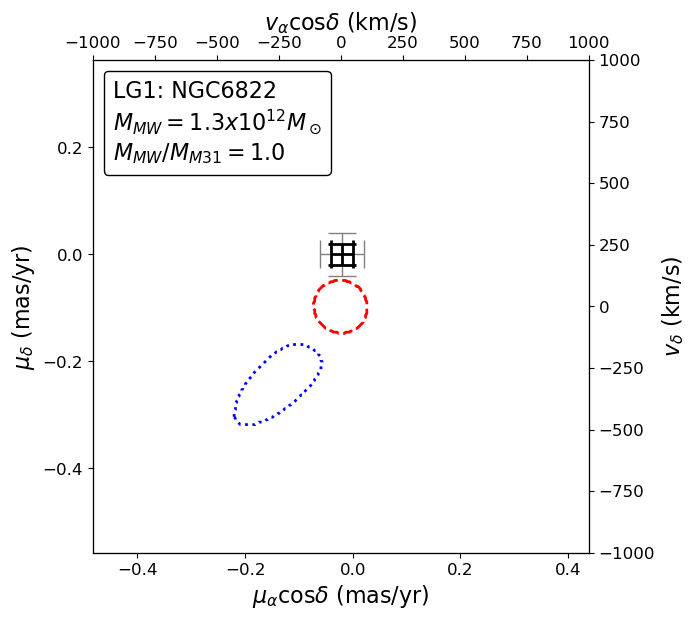}
    \includegraphics[width=0.3\textwidth]{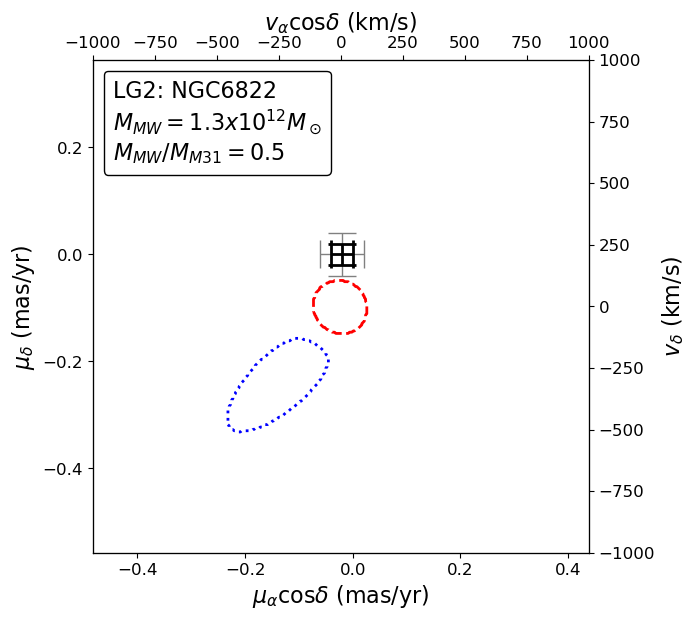}
    \includegraphics[width=0.3\textwidth]{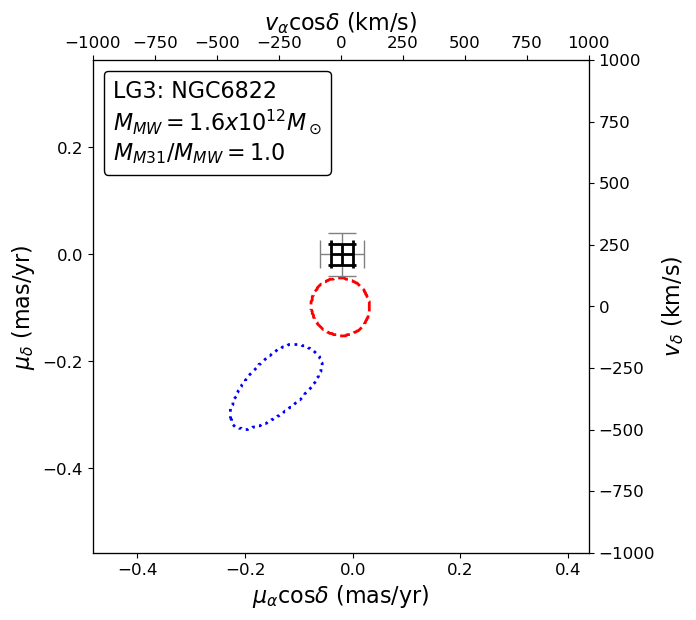}
    \caption{Results of the search through orbital parameter space for
      Eridanus 2, Phoenix, Leo T and NGC 6822. The different panels for each dwarf
correspond to the LG1, LG2 and LG3 setups (left to right,
respectively; see text for details). Red dashed lines delineate the regions of parameter space
where the orbits bring the dwarf within 300\,kpc of the Milky
Way. Blue dotted lines delineate the regions of parameter space
where the orbits bring the dwarf within 300\,kpc of M31. If no dotted or
dashed lines are present, it means there is no region of parameter
space that brings the dwarf within 300\,kpc of these galaxies. Overlaid on each panel is the measured proper motion of the
dwarf as given in Table~\ref{pmresults}. Black lines show random
errors, and grey lines also include the systematic uncertainties.}\label{pmspace1}
\end{figure*}

\begin{figure*}
    \includegraphics[width=0.3\textwidth]{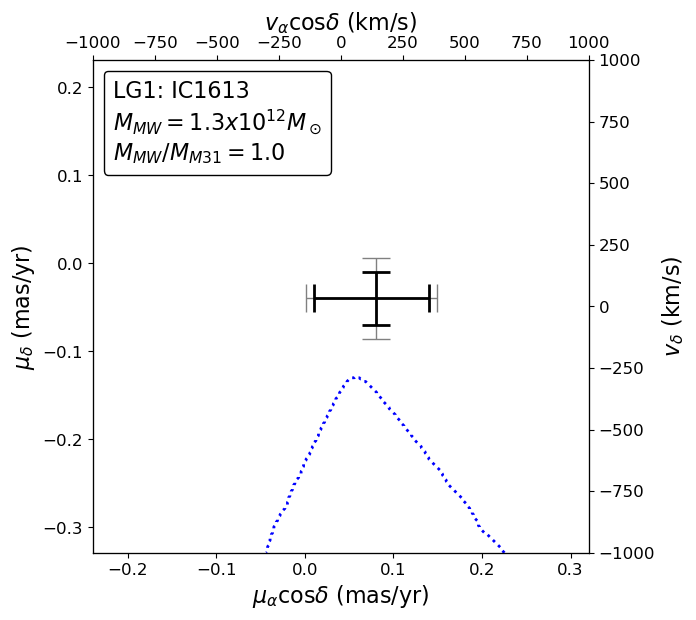}
    \includegraphics[width=0.3\textwidth]{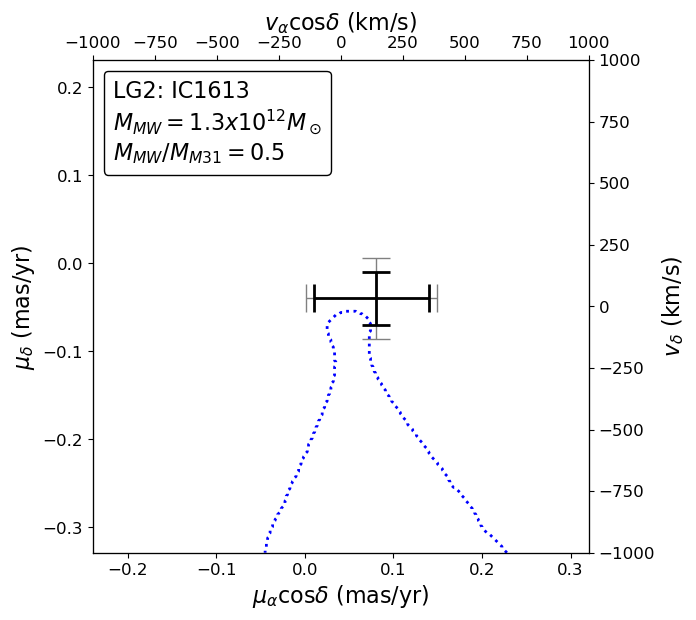}
    \includegraphics[width=0.3\textwidth]{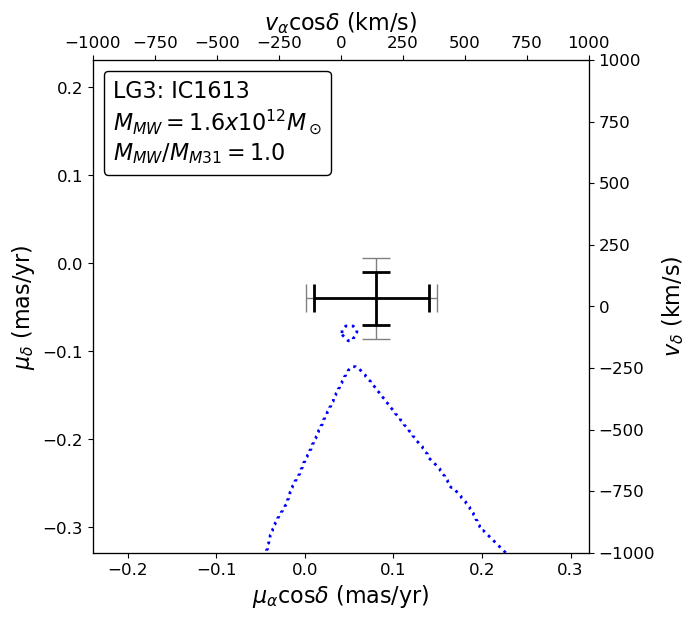}
    \includegraphics[width=0.3\textwidth]{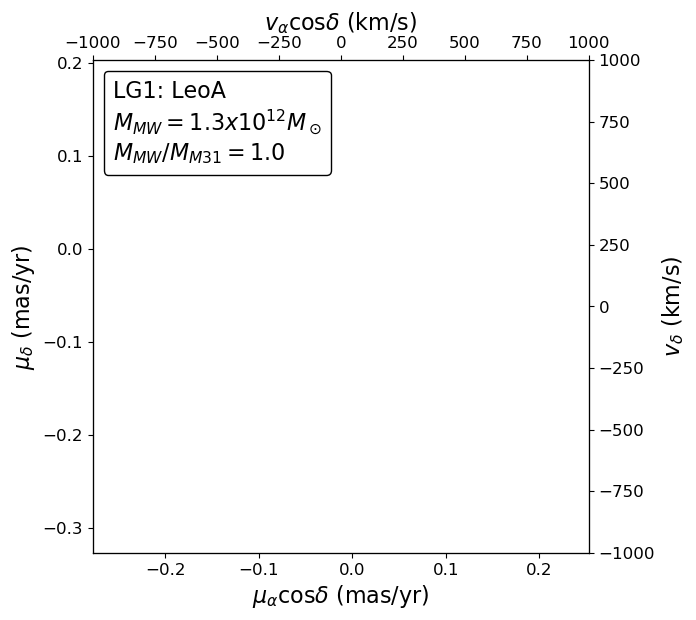}
    \includegraphics[width=0.3\textwidth]{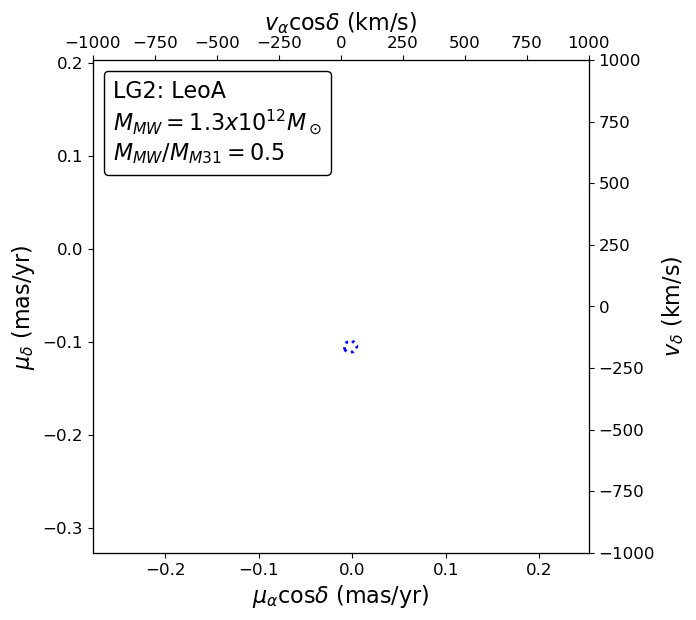}
    \includegraphics[width=0.3\textwidth]{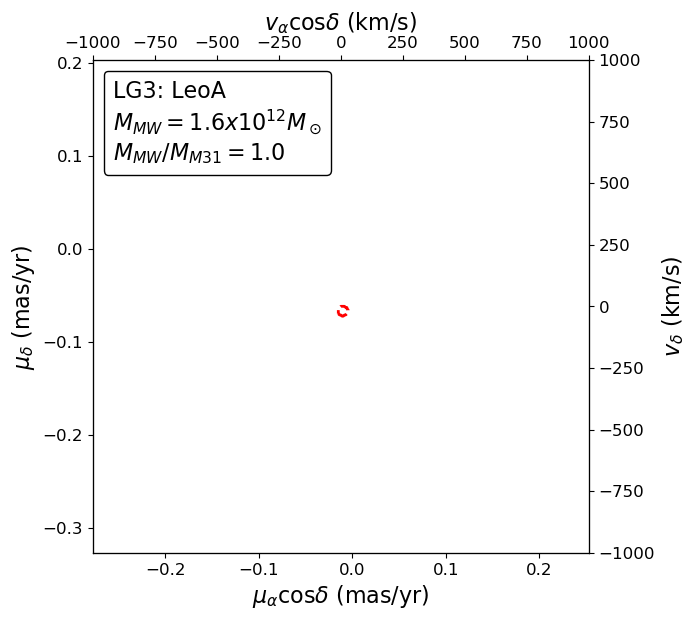}
    \includegraphics[width=0.3\textwidth]{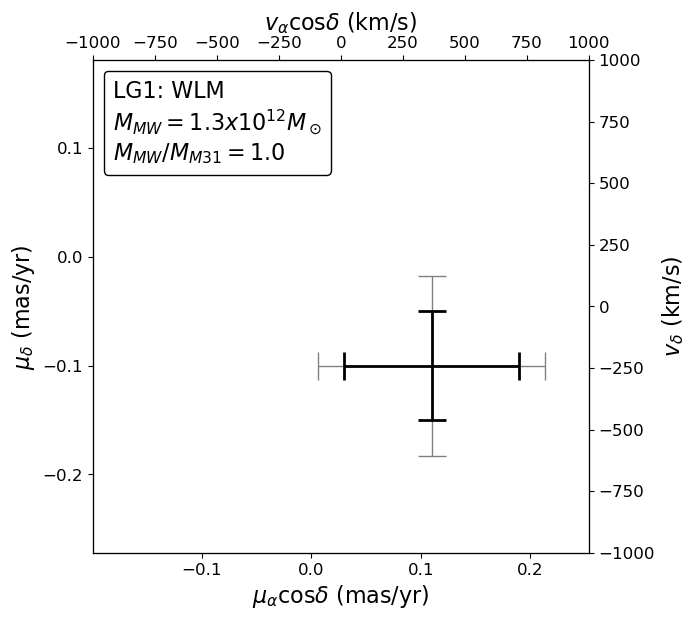}
    \includegraphics[width=0.3\textwidth]{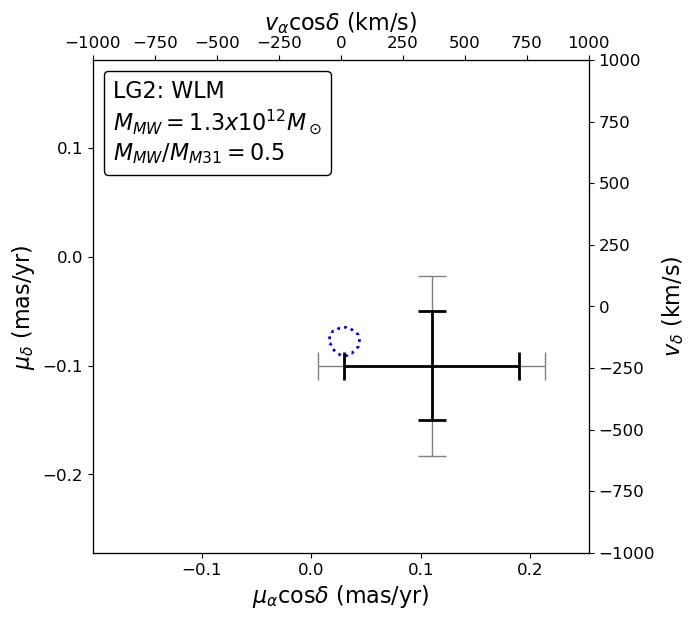}
    \includegraphics[width=0.3\textwidth]{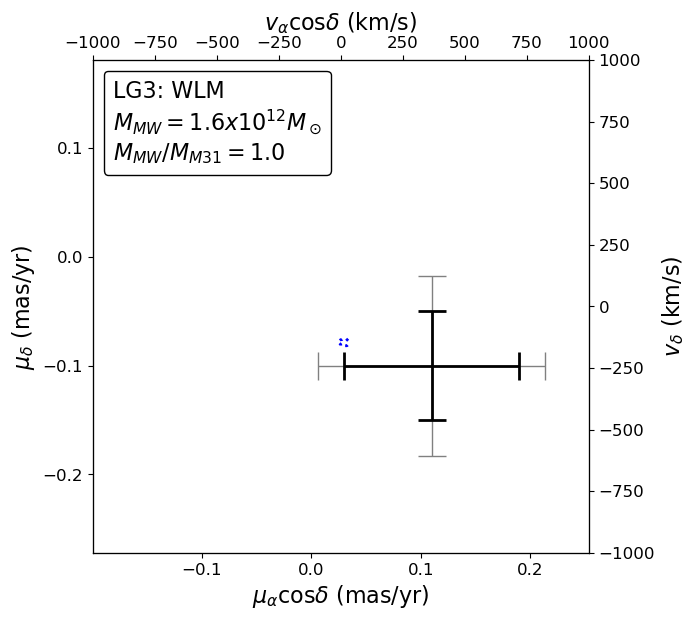}
    \caption{Same as Figure~\ref{pmspace1}, for IC 1613, Leo A and
      WLM. Note that the measured systemic proper motion of Leo A
      falls outside of the proper motion space explored (although it is
      consistent with zero at less than 2\,$\sigma$).}\label{pmspace2}
\end{figure*}

Figures~\ref{pmspace1} and \ref{pmspace2} show the results of these
simulations for each dwarf. The different panels for each dwarf
correspond to the LG1, LG2 and LG3 setups (left to right,
respectively). Red dashed lines delineate the regions of parameter space
where the orbits bring the dwarf within 300\,kpc of the Milky
Way. Blue dotted lines delineate the regions of parameter space
where the orbits bring the dwarf within 300\,kpc of M31. If no dotted or
dashed lines are present, there is no region of parameter
space that brings the dwarf within 300\,kpc of these galaxies. Overlaid on each panel is the measured proper motion of the
dwarf as given in Table~\ref{pmresults}. Black lines show random
errors, and the grey lines also include the systematic uncertainties
described earlier.


For the 5 galaxies that are currently closer to the Milky Way than
M31 - Eridanus 2, Phoenix, Leo T, NGC 6822 and Leo A - there is a region of parameter space centered around a tangential
velocity of zero in which these galaxies
could have had an interaction with the Milky Way. However, in the
cases of Phoenix and Leo A, these only exist for a more massive Milky
Way. Even here, any tangential velocity in excess of $35 - 40$\,km\,s$^{-1}$ or so for each dwarf would move the
galaxy into the family of orbits which have not had an interaction.

For WLM and IC 1613, no solutions are found in which they have
interacted with the Milky Way. For WLM, it is only possible for it to
have interacted with M31 in the scenarios where M31 is relatively
massive. There is a reasonably large region of parameter space in
which IC 1613 could have interacted with M31.

It is interesting to note that NGC 6822, Leo T and Leo A could have
interacted with either the Milky Way or M31, depending on their
tangential velocities. For the latter two, M31 interactions can only
have occured in a relatively small region of parameter space for the
setup with the most massive M31 ($\sim 2.6 \times 10^{12}
M_\odot$). For NGC 6822, there is a reasonably large area of parameter
space for M31 or Milky Way interactions for all three setups.

Figures~\ref{pmspace1} and \ref{pmspace2} provide a simple context
to understand the usefulness, or otherwise, of
the current proper motion measurements for each of the dwarfs:

\begin{itemize}
\item {\bf Eridanus 2:} This dwarf can only have passed within 300\,kpc 
  of the Milky Way if its Galactocentric tangential velocity is less
  than $\sim 90$\,km\,s$^{-1}$. This is at the $1-\sigma$ edge of the current
  proper motion measurement. The allowable parameter space for a
  backsplash solution is consistent with a similar study by
  \cite{blana2020}, who consider the orbits of Eridanus 2, Phoenix,
  Leo T and Cetus;
\item {\bf Phoenix:} Of the explored setups, the only way for Phoenix
  to have ever passed with 300\,kpc of the Milky Way if it has
 a small ($\lesssim 35$\,km\,s$^{-1}$) tangential velocity and the total mass of the Milky
  Way is relatively massive (as in LG3; at the upper end of the mass range to
  which it has been constrained by recent measurements). This is
  also consistent with the study by \cite{blana2020}. If the Milky
  Way is really this massive, then the
  uncertainties on the current proper motion are currently a factor of
  a few too large to distinguish a backsplash scenario from a more
  isolated orbit. As Phoenix requires the Milky Way to be massive in
  order for an interaction to occur, tightening constraints on the
  Milky Way's total mass will likely prove valuble in definitely ruling out a backsplash origin for Phoenix;
\item {\bf Leo T:} If Leo T has a tangential velocity less than $\sim
  80$\,km\,s$^{-1}$, then it is probably a Milky Way backsplash system, as also concluded
  by \cite{blana2020}. Interestingly, in the case of a very massive
  M31 (as in LG2), there is also a small region of parameter space in
  which Leo T could be a M31 backsplash system (this is also the case
  for Leo A; see later). However, the current
  proper motion measurement is utterly useless for providing any
  meaningful constraints in this respect. It is worth noting that the
  member stars of Leo
  T (and to a lesser extent, Leo A) have not been observed as
  frequently as in other dwarf galaxies as a result of the Gaia scanning law. In particular, the average number
  of ``along-scan'' measurements by the spacecraft for the member stars of Leo T 
  is {\tt astrometric\_n\_obs\_al} $\simeq 100$ ({\tt
    astrometric\_n\_obs\_al} $\simeq 127$ for Leo A). For comparison,
  NGC 6822 and IC 1613 have around
  50\% more measurements on average ($\sim 150 - 170$), WLM and Eridanus 2 have
  about twice as many ($\sim 200 - 240$), and Phoenix has more than 3
  times as many (more than 300, on average);
\item {\bf NGC 6822:} In principle, NGC 6822 could have
  had interactions with either the Milky Way or M31, but the relatively
  precise proper motion measurement means that an M31 interaction is
  unlikely at more than $4 - \sigma$. A Milky Way
  interaction requires NGC 6822 to have a Galactocentric tangential
  velocity of less than $\sim 110$\,km\,s$^{-1}$, and this is at the edge of the
  $1-\sigma$ proper motion uncertainties. It is worth noting the
  importance of systematic errors for NGC 6822, since these contribute
  more than random uncertainties. We examine the orbit of NGC 6822 in
  more detail in the next section;
\item {\bf IC 1613:} The current proper motion measurement of IC 1613 is
  inconsistent at approximately $2-\sigma$ with ever having an
  interaction with M31, under the assumption that M31's mass is only
  around $1.3 \times 10^{12}M_\odot$ (as in LG1). The region of parameter space in
  which an interaction can occur corresponds broadly to IC 1613's
  current space
  velocity pointing away from M31, and aligned within $\sim 40$ degrees of the radial vector
  from M31. For a more massive M31 (as in LG2 and LG3),
  then the proper motion of IC 1613 more closely overlaps with the regime of
  parameter space in which interactions with M31 can occur;
\item {\bf Leo A:} There is only a tiny regime of parameter space in
  which Leo A could have interacted with either the Milky Way or
  M31. For the latter, M31 must be massive (as in LG2; around twice the most
  likely mass for the Milky Way). For the former, the Milky Way must
  also be relatively massive (as in LG3; at the upper end of the current
  preferred mass range) and the relative tangential velocity of the
  two must be less than around 40\,km\,s$^{-1}$. Like Leo T, the current estimate for the
  proper motion of Leo A is completely useless for distinguishing
  these scenarios, but the relatively tiny regions of parameter space
  that need to be excluded means that future improvements in the size
  of the uncertainities associated with the proper motion may put
  meaningful constraints on the likelihood or otherwise of a backspash origin;x
\item {\bf WLM:} Within the setups explored, this dwarf can only have
  interacted with M31, and can only have done so if M31 is relatively
  massive (as in LG2 and LG3). Those orbits which lead to an interaction require WLM to be
  close to apocenter, and with relative velocities between the two
  less than $\sim 60$\,km\,s$^{-1}$. This region of parameter space
  overlaps with our current estimate of its proper motion.
  \end{itemize}

\subsection{The orbit of NGC 6822}

\begin{figure*}
    \includegraphics[width=0.8\textwidth]{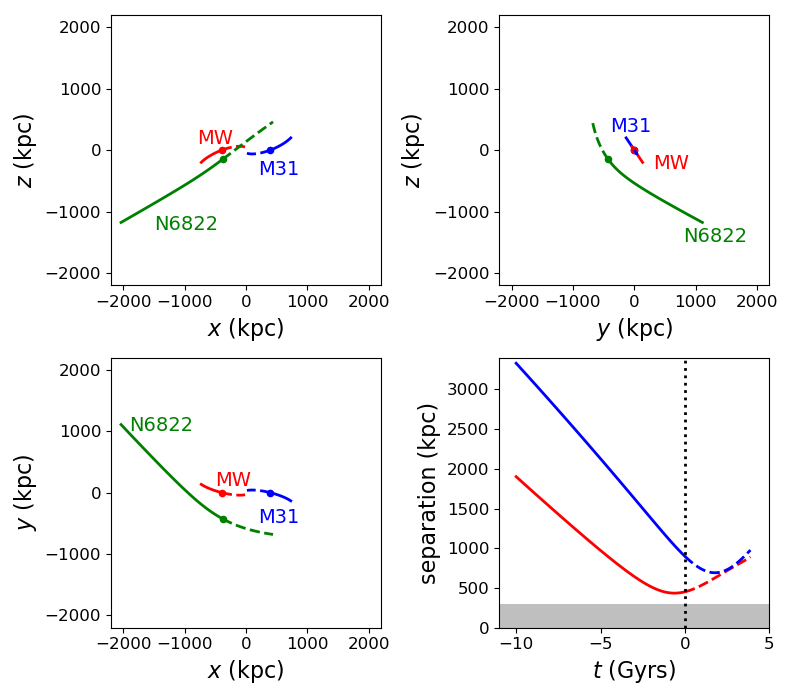}
    \caption{First three panels show the projection of the most likely orbit of NGC 6822 (green)
      relative to the Milky Way (red) and M31 (blue) over the last
      10\,Gyrs (solid lines), and for the next 4\,Gyrs (dashed lines),
      using the LG1 potential. Solid
      points indicate current positions of the galaxies. The
      coordinate system is centered on the mid point of the Milky Way
      - M31 positions, and aligned so that the x-axis connects the
      Milky Way and M31. The fourth panel shows the separation of the
      dwarf as a function of time from the Milky Way (red line) and
      M31 (blue line). Present day is indicated by the dotted
      line. The grey shaded region corresponds to a separation of less
    than 300\,kpc from either large galaxy}\label{6822orbit}
\end{figure*}

While several of the galaxies listed in Table~\ref{pmresults} have
good proper motion measurements in an angular sense (better than a few
tenths of a mas/yr), only NGC 6822 has
a tangential velocity constrained to much better than 100\,km\,s$^{-1}$ or so
(even after systematic uncertainities are taken into account). Thus, compared to any of the other galaxies in
Figures~\ref{pmspace1} and \ref{pmspace2},  the volume of likely parameter space
allowed by the proper motion measurement of NGC 6822  is considerably smaller than
the volume of parameter space probed by our previous grid search. As
such, we decided to conduct a more detailed analysis of the possible
orbits of NGC 6822 within the Local Group.

Figure~\ref{6822orbit} shows the orbital path of NGC 6822 in the Local
Group over the last 10\,Gyrs, for the LG1 realisation. The first three
panels show projections in the x-y-z space, and the fourth panel shows
the separation of the dwarf as a function of time from the Milky Way
and M31 (red and blue lines, respectively). Solid points indicate the current
positions of the three galaxies. The coordinates are based on the
Galactocentric frame, but rotated so that M31 lies on the x axis, and
centered on the midpoint
of the Milky Way - M31 positions. Here,
we have set the radial velocity, proper motion and distance of M31 to
the values used earlier. We have also set the radial velocity and distance
of NGC 6822 to the values used earlier, and we have set the proper
motion of NGC 6822 to the measured value. In this idealised scenario,
NGC 6822 is clearly on its first infall into the Local Group and has had no
interaction with either of the big galaxies. For most of its history,
it has been falling nearly radially into the Local Group, and 10\,Gyrs
ago it was approximately 1.9\,Mpc away from the Milky Way (3.3\,Mpc away
from M31).

To better understand the range of orbits consistent with the
observational parameters, we use the three Local Group setups, with the different values of
$M_{MW}$ and $f$, described previously. For each setup, 
$10^5$ realisations of the Milky Way, M31 and NGC 6822 are evolved backwards in
time for 10\,Gyrs. For each realisation, values for the distance modulus,
the heliocentric radial velocity and the proper motion of both
NGC 6822 and M31 are selected from Gaussian distributions. These are centered on the relevant measured
values and have a standard deviation equal to the uncertainty on the
measured value.

Out of the three sets of $10^5$ realisations, only eight orbits are
found that involve NGC 6822 passing within
300\,kpc of M31, in line with our findings from the grid search, and
essentially ruling out a past M31 interaction for
NGC 6822. In constrast, $6 - 9$\% of the realisations involve a
$\leq 300$\,kpc passage
between NGC 6822 and the Milky Way. This too is broadly in line
with the results from the grid search, which did not take into account
the uncertainties in all the other measured parameters for
NGC 6822 and M31.

More specifically, for each set of $10^5$ realisations:

\begin{enumerate}
  \item LG1: $6631 \pm 81$ orbits ($6.6\%$) pass within 300\,kpc of the
    Milky Way; 1 orbit
    passes within 300\,kpc of the M31 (uncertainties are Poisson statistics); 
  \item LG2: $6491 \pm 81$ orbits ($6.5\%$) pass within 300\,kpc of the Milky Way; 6 orbits
    pass within 300\,kpc of the M31; 
  \item LG3: $8298 \pm 91$ orbits ($8.3\%$) pass within 300\,kpc of the Milky Way; 1 orbit
    passes within 300\,kpc of the M31.
\end{enumerate}

We conclude from this analysis that the current uncertainties on the
relevant observational parameters imply that NGC 6822 has not had any
interactions with M31 or the Milky Way, at $\sim 99.99\%$
confidence and better than 90\% confidence, respectively. We also note that the
mass of M31 does not have any significant impact on the likelihood
that NGC 6822 is a backsplash galaxy or not.

\begin{figure*}
    \includegraphics[width=0.8\textwidth]{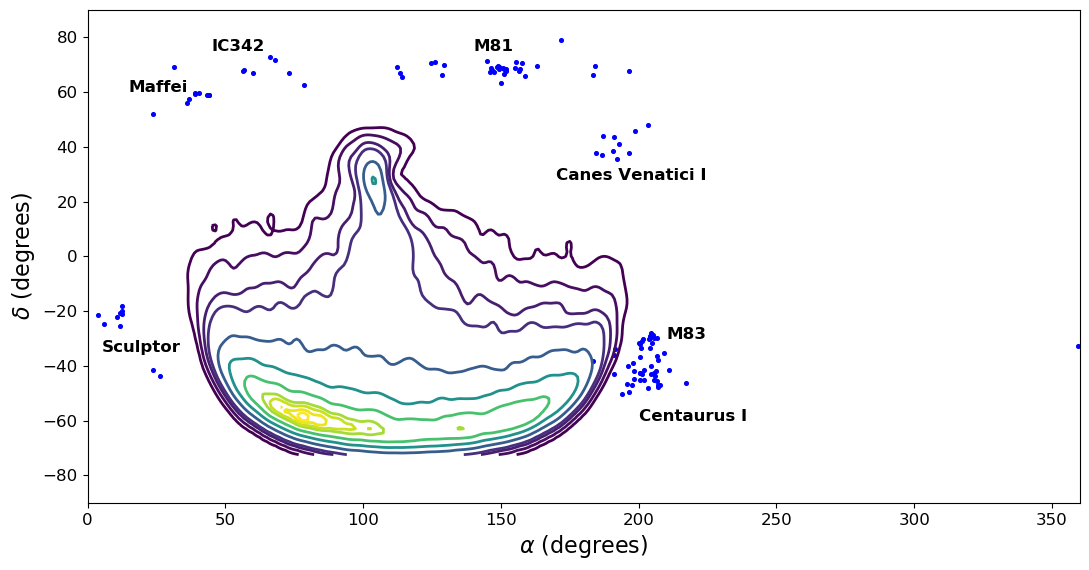}
    \caption{Contour map of the predicted positions of NGC 6822 10\,Gyrs
      ago, in celestial coordinates, for those orbits in which NGC
      6822 did not pass within 300\,kpc of the Milky Way, in the LG1
      realisation. $2 \times 2$\,degree pixels were used, and the
      counts per pixel were weighted by $\cos\delta$. Counts were
      smoothed by a Gaussian with $\sigma = 1$ pixel. Contour levels
      are set at 2, 5, 10, 30, 70, 80, 95 and 98\% of the maximum
      value. Also shown are the (current) locations of galaxies in
      nearby groups from Karachentsev et al. (2005).}\label{origins}
\end{figure*}

Given that NGC 6822 appears to be on its first infall into the Local
Group, a natural question to ask is ``infall from
where?''. Figure~\ref{origins} shows a contour map of the predicted
positions of NGC 6822 10\,Gyrs ago, in celestial coordinates, for
those orbits in which NGC 6822 did not pass within 300\,kpc of the Milky
Way. The LG1 realisation was used to make this figure, although all
three realisations are qualitatively similar. Further, the
distribution of points is similar whether the plot is made for
10\,Gyrs ago or 8\,Gyrs ago, as is expected given that at these early
times NGC 6822 is far from the Galaxy and falling mostly radially
towards it. $2 \times 2$\,degree pixels were used, and the counts per
pixel were weighted by $\cos\delta$. Counts were smoothed by a
Gaussian with $\sigma = 1$ pixel. Contour levels are set at 2, 5, 10,
30, 70, 80, 95 and 98\% of the maximum value.

Also shown in Figure~\ref{origins} are the (current) locations of galaxies in nearby
groups from \cite{karachentsev2005}. The closest of these groups (Maffei) is
approximately 3\,Mpc away, and their projected  positions on the sky should
not have shifted significantly over the last 10\,Gyrs given that mostly
radial motion relative to the Local Group is expected. Note that the
very nearby NGC3109 galaxy group is not shown. Its very close
proximity to the Local Group means that it is quite possible that it
and the Local Group have undergone a more complex, non-radial
orbit. This means that the present location of this galaxy group
($\alpha \sim 151^\circ, \delta \sim -26^\circ$) on the sky is
not necessarily a good indicator of its historic position. It is well
known that all the closest galaxy groups are found along a great
circle path, indicative of the Local Group being embedded in a ``local sheet''
of nearby structures (e.g., \citealt{tully2008} and references therein).

There is clearly a reasonably-well defined locus on the sky from which
NGC 6822 likely originated. This locus is assiduous in avoiding
any overlap with the locations of any of the nearest galaxy
groups. This is perhaps to be expected: should NGC 6822 have originated
in a dark matter halo in the vicinity of any of the other galaxy
groups, then it would likely have ended up as a satellite in one of
these groups. So this appears consistent with the idea that NGC 6822
formed in relative isolation in the periphery of the structure now
known as the Local Group of galaxies, close enough that the gravity of
the Local Group won out over the Hubble expansion. Over the course of
a Hubble time, it has fallen in to the Local Group, and is now in
relatively close proximity to the Milky Way. We note that there have
been suggestions that NGC6822 may have undergone a recent interaction
with another small stellar system (e.g., see
\citealt{deblok2000, demers2006}, but see also \citealt{cannon2012, thompson2016}), in which case perhaps NGC
6822 had a dwarf-like companion at earlier times.

And what of the immediate future of NGC 6822? Integrating forward the orbital
realisations from LG1 for which there was not a previous interaction suggests that it has only a 0.04/0.6\% chance of
having an interaction (i.e., passing within $\sim 300$\,kpc) with the
Milky Way/M31 in the next 4\,Gyrs. But this uneventful future for
NGC 6822 is not shared by the Milky Way and M31. For in about 4\,Gyrs,
these two large galaxies will be within about 100\,kpc of each other,
in the midst of a near head-on collision (e.g., see
\citealt{vandermarel2012b}). The most likely fate for
NGC 6822 over the next 4\,Gyrs is shown with the dotted line in
Figure~\ref{6822orbit}. In this realisation, NGC 6822 is currently
near pericenter with the Milky Way, and will continue on its
relatively uneventful path through space, oblivious to the imminent
formation of the massive galaxy formerly known as the Local Group.

\section{Discussion and summary}

\subsection{Proper motion measurements and future Gaia data releases}

The
parameter space probed in Figures~\ref{pmspace1} and \ref{pmspace2}
gives a good handle on the current usefulness of the orbital
constraints for the dwarf galaxies
provided by the new measurements of their proper motions. From these considerations, and
our orbital analysis, we
conclude that the proper motion measurement of
NGC 6822 in Table~\ref{pmresults} is interesting, even including
systematic uncertainties. IC 1613 and WLM both have good measurements
of their proper motions in an angular sense, but given their distances, their tangential velocities are less
interesting. The current proper motion measurement for  the remaining galaxies are not
particularly useful. While smaller uncertainties are needed, we note
that the currently derived proper motions are
generally within $2-\sigma$ of the predicted proper motions by
\cite{shaya2013}.

It is important to recognise the impact future Gaia data releases will
have on these measurements. Firstly, it is unlikely that any of the
galaxies listed in Table~\ref{gals} but not in Table~\ref{pmresults}
will ever have any meaningful measurements of their proper motion
using Gaia data: there are simply not enough bright stars visible to
Gaia in these galaxies. Secondly, random errors in all these
measurements will decrease significantly as the baseline ($t$) of the
Gaia observations increases and the signal to noise of the
observations increases, leading to a proportionality of
$t^{-3/2}$. For example, the imminent release of Gaia Data Release 3
in December 2020 will be based on 34 months of observations compared
to 22 months for Data Release 2, and proper motions are expected to be
a factor of 2 better than
DR2\footnote{\url{https://www.cosmos.esa.int/web/gaia/earlydr3}}. This
should mean that the tangential velocities of IC 1613, WLM and possibly
Phoenix will be able to be constrained at an interesting level,
especially if systematic uncertainties also decrease. By the end of
the complete mission in 2022, with an 8 year baseline, it seems likely
that the random uncertainties on the tangential velocities should
decrease by a factor of 9 in comparison to these DR2 results. In this
case, all the galaxies in Table~\ref{pmresults} except possibly Leo T
and Leo A should have uncertainties in their tangential velocities of
several tens of km\,s$^{-1}$ in each direction or better. By this point, the
velocity components of NGC 6822 should be measured with exquisite
precision.

\subsection{Non-Gaia proper motions}

So what of Leo T, Leo A, and the {\it Solo} dwarfs listed in
Table~\ref{gals} that do not have (m)any stars visible in Gaia? More
sensitive observatories are clearly required to detect a sufficient
number of their stars,
but combining sensitivity with high spatial resolution and low
systematic errrors means there are only a few facilities in the
optical and IR regimes that could make these measurements in the next decade.

There are many large aperture facilities on the ground. Most notably,
the Legacy Survey for Space and Time, to be 
conducted by the Vera C. Rubin Observatory, will be conducted over a
period of a decade. Its repeated observations of the entire
southern hemisphere will make it very suitable for proper motion
studies, and its data acquisition strategy and heavy focus on data
processing will likely ensure a good understanding of systematic
errors. However, as a seeing-limited facility, its spatial resolution
will likely not generally be sufficient for many proper motion studies
beyond the Milky Way sub-group. In comparison, the very high spatial resolution of the Thirty Meter
Telescope (TMT) and the Extremely Large Telescope (ELT) equipped with
multi-conjugate adaptive optics systems, will certainly provide very
high spatial resolution (e.g., see \citealt{evslin2015} for the
science that can be enabled). Here, however, the challenge will be to
obtain a
sufficiently detailed understanding and characterisation of the 
systematic uncertainties which will otherwise dominate, and which are
contributed to by residual spatial distortions from the adaptive optics
system (e.g., see
Taheri et al., 2020, submitted, and references therein).

Space-based observatories will therefore provide the most promising
avenue for more Local Group proper motions.  It is worth highlighting
the proven utility of HST for these types of measurements. HST has
been used to conduct proper motion studies of dwarf galaxies for many
years, with early ground-breaking work that utilised background
quasars to define the absolute reference frame (e.g.,
\citealt{piatek2002, piatek2003, piatek2005, piatek2006, piatek2007,
  kallivayalil2006a, kallivayalil2006b, kallivayalil2013}). Later
adoption of a reference frame defined by background galaxies reduced
the restriction in field selection caused by the requirement for there
to be a background quasar. Since then, proper motion measurements
beyond the Milky Way subgroup have been made for M31
(\citealt{sohn2012}), NGC147 and NGC185 (\citealt{sohn2020}). Proper
motion measurements also exist for M33 and IC10 from VLBI
observations (\citealt{brunthaler2005, brunthaler2007}). We note that
Gaia DR2 has also been used to independently obtain the proper motions
of M31 and M33 (\citealt{vandermarel2019}), although at these distances the Gaia DR2 accuracy is
significantly less than for HST (and, of course, VLBI).

Upcoming space missions in the next decade with the potential to
provide new Local Group proper motion measurements include the  James Webb Space Telescope (JWST), the Nancy
Grace Roman Space Telescope, and the Euclid space mission. These observatories will all be active by mid-decade, and all operate at longer
wavelengths than HST.  Taking into account their apertures, their
spatial resolutions will either be similar or slightly poorer than HST
(but still very good in an absolute sense). However, a major advantage of HST
for proper motion studies is the very long baseline of observations
that exist given the maturity of this facility. In this regard, it is
very likely that, even using data on these dwarfs from JWST, Roman or Euclid, first
epoch observations will be from HST. Indeed, programs are already
underway with HST to obtain first epoch observations of Local Group
dwarf galaxies where none currently exist, or to obtain second-epoch
observations where first-epoch already exist. Significant survey datasets in
this regard include the {\it Advanced Camera for Surveys Local Cosmology from Isolated Dwarfs} (ACS LCID) survey for isolated dwarfs
(e.g., \citealt{cole2007, monelli2010a, monelli2010b, hidalgo2011,
  skillman2014}), the
{\it Initial Star formation and Lifetimes of Andromeda Satellites} (ISLANDS) for M31 dwarf spheroidals (\citealt{monelli2016, skillman2017,
martinezvazquez2017}), and work on fainter
members of the M31 satellite system (\citealt{martin2017, weisz2019}).

\subsection{Summary}

 In this paper, we  have measured systemic proper motions for distant ($d \sim 450 - 900$\,kpc) dwarf galaxies in the Local
  Group using CFHT and Gaia Data Release 2 and investigated if these
  (currently) isolated galaxies have ever had an interaction with either the Milky
  Way or M31. We find that NGC 6822, Leo A,
  IC 1613 and WLM have a sufficient number of bright stars with reliable
  astrometry to derive proper motions for these four
  galaxies. We apply a variant of the methodology described in MV2020
  to obtain estimates of their systemic proper motions. The results for
  NGC 6822, IC 1613 and WLM are good in an angular sense - systematic
  errors in Gaia DR2 proper motions either dominate or are major
  contributers to the error budgets. However, the large distances of
  IC 1613 and WLM mean that the resulting constraint on their orbits are still
  relatively weak.

  We explore the orbital parameter space for these isolated galaxies
  in light of these new measurement, and also include Leo T, Eridanus
  2 and Phoenix, which have recent proper motion estimates from
  MV2020. We conclude that Phoenix, Leo A and WLM are unlikely to have
  ever interacted with the Milky Way or M31, unless these galaxies are
  very massive (at the upper end of any recent estimates of their
  masses). We rule out a past interaction of NGC 6822 with M31 at
  $\sim 99.99\%$ confidence, and find there is less than a 10\% chance
  that NGC 6822 has had an interaction with the Milky Way. We also
  explore the origins of this galaxy in the periphery of the young
  Local Group. For the
  remaining dwarfs under consideration, the current proper motions do not yet rule out the possibility
  that these galaxies may be backsplash systems. However, our results
  indicate that future data releases from Gaia will provide good or
  excellent constraints on the interaction history of these seven
  galaxies, with the possible exceptions of Leo A and Leo T. These
  last two galaxies suffer from a combination of only a few bright
  stars being visible to Gaia, and a relative dearth of visits by the
  spacecraft as a result of the Gaia scanning law.

\section*{Acknowledgements}

We thank the anonymous referee for comments that improved the
paper. We also thank Mat{\'i}as Bla{\~n}a and Tony Sohn for their comments.

AWM and KAV would like to acknowledge funding from the National Science and Engineering Research Council 
Discovery Grants program. GB acknowledges financial support through
the grants (AEI/FEDER, UE) AYA2017-89076-P, as well as by the
Ministerio de Ciencia, Innovación y Universidades (MCIU), through the
State Budget and by the Consejer a de Economia, Industria, Comercio y
Conocimiento of the Canary Islands Autonomous Community, through the
Regional Budget.

Based on observations obtained with MegaPrime/MegaCam, a joint project
of CFHT and CEA/DAPNIA, at the Canada-France-Hawaii Telescope (CFHT)
which is operated by the National Research Council (NRC) of Canada,
the Institut National des Science de l'Univers of the Centre National
de la Recherche Scientifique (CNRS) of France, and the University of
Hawaii. The observations at the Canada-France-Hawaii Telescope were
performed with care and respect from the summit of Maunakea which is a
significant cultural and historic site.

\section*{Data availability}

Some of the data underlying this article were accessed from the Gaia
Archive (\url{https://gea.esac.esa.int/archive/}). The CFHT data
underlying this article are available through the CFHT archive hosted
by the Canadian Astronomy Data Center
(\url{http://www.cadc-ccda.hia-iha.nrc-cnrc.gc.ca/en/cfht/}). Derived
data products will be shared on reasonable request to the corresponding author.


\bibliographystyle{mnras}


\end{document}